\def\TS{T}

\if T\TS
  \input psbox.sty
  \documentstyle[aps,prb,twocolumn,jpc,print2]{revtex}
\else
  \documentstyle[aps,prb,preprint,jpc]{revtex}
  
\fi

\def\FIGone{
\begin{figure}[htbp]
\centerline{\psboxto(5cm;0pt){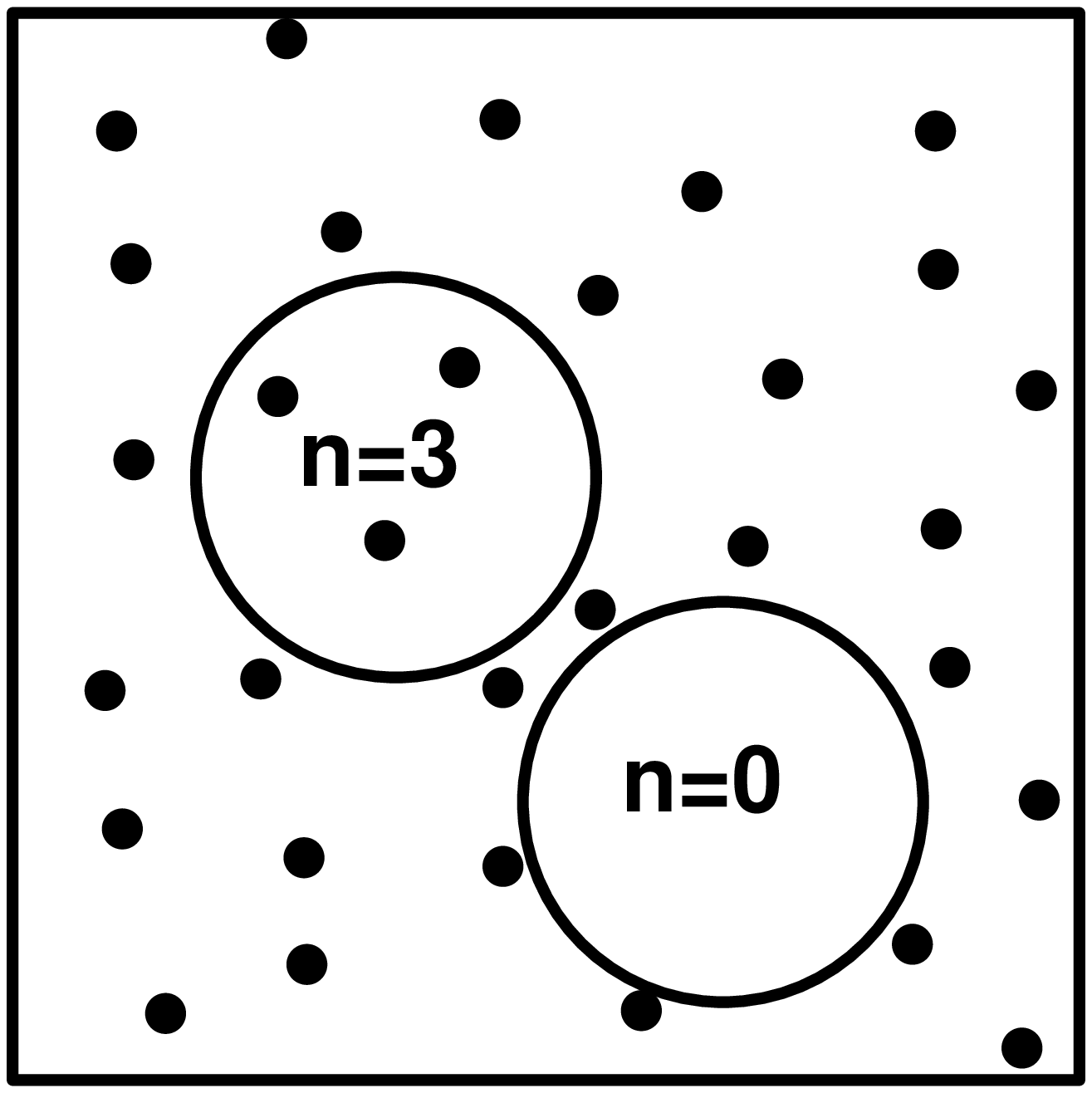}}
  \caption{Schematic two-dimensional representation of the
    probabilities $p_n$ of observing $n$ solvent centers inside a
    randomly positioned volume $v$ in the bulk fluid.  Two cases are
    shown: A successful insertion with $n=0$ that would contribute to
    $p_0$, and an insertion with $n=3$ contributing to $p_3$.}
  \label{fig:pzero}
\end{figure}
}

\def\FIGtwo{
\begin{figure}[htbp]
\centerline{\psboxto(7cm;0pt){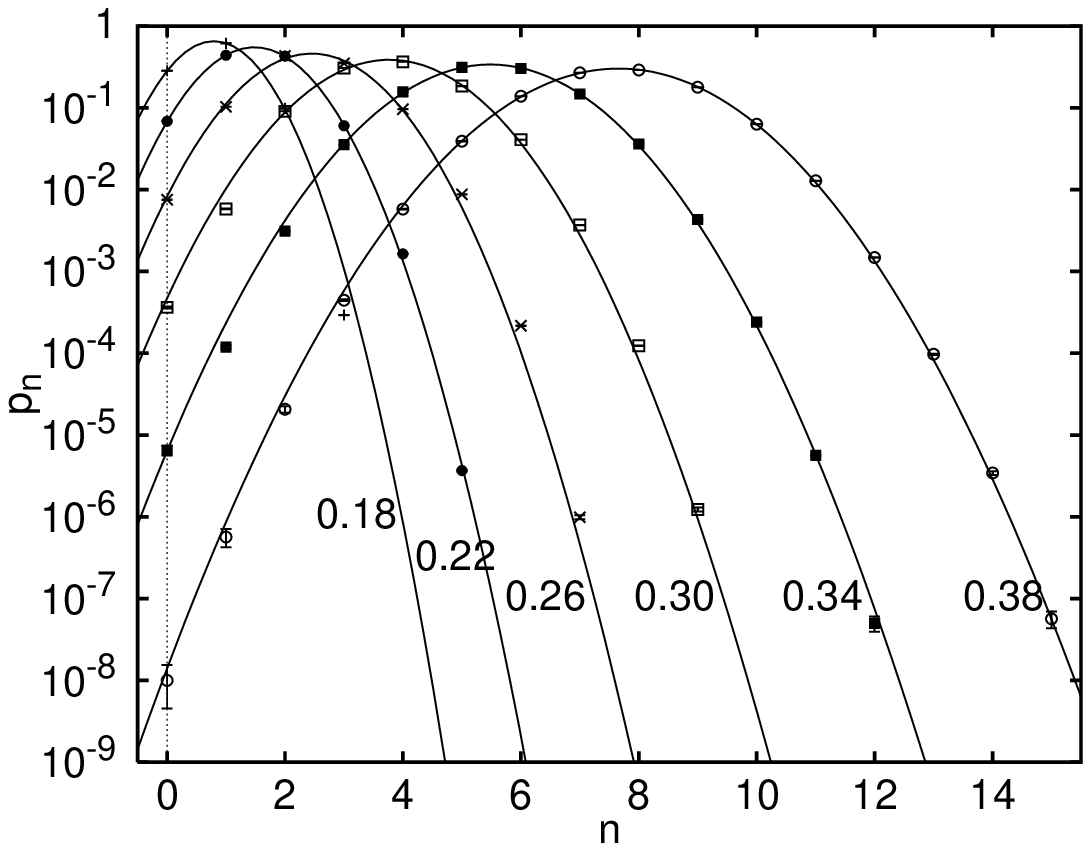}}
  \caption{Probabilities $p_n$ for observing $n$ water-oxygen atoms in
    spherical cavity volumes.\protect\cite{Hummer:PNAS:96} Results
    from Monte Carlo simulations of SPC
    water\protect\cite{Berendsen:81} are shown as symbols.  The
    parabolas are the predictions of the flat default model of the
    IT approach.  The center-to-center exclusion
    distance $d$ (in nanometers) is noted next to the curves.}
  \label{fig:sphere_pn}
\end{figure}
}

\def\FIGthree{
\begin{figure}[htbp]
\centerline{\psboxto(7cm;0pt){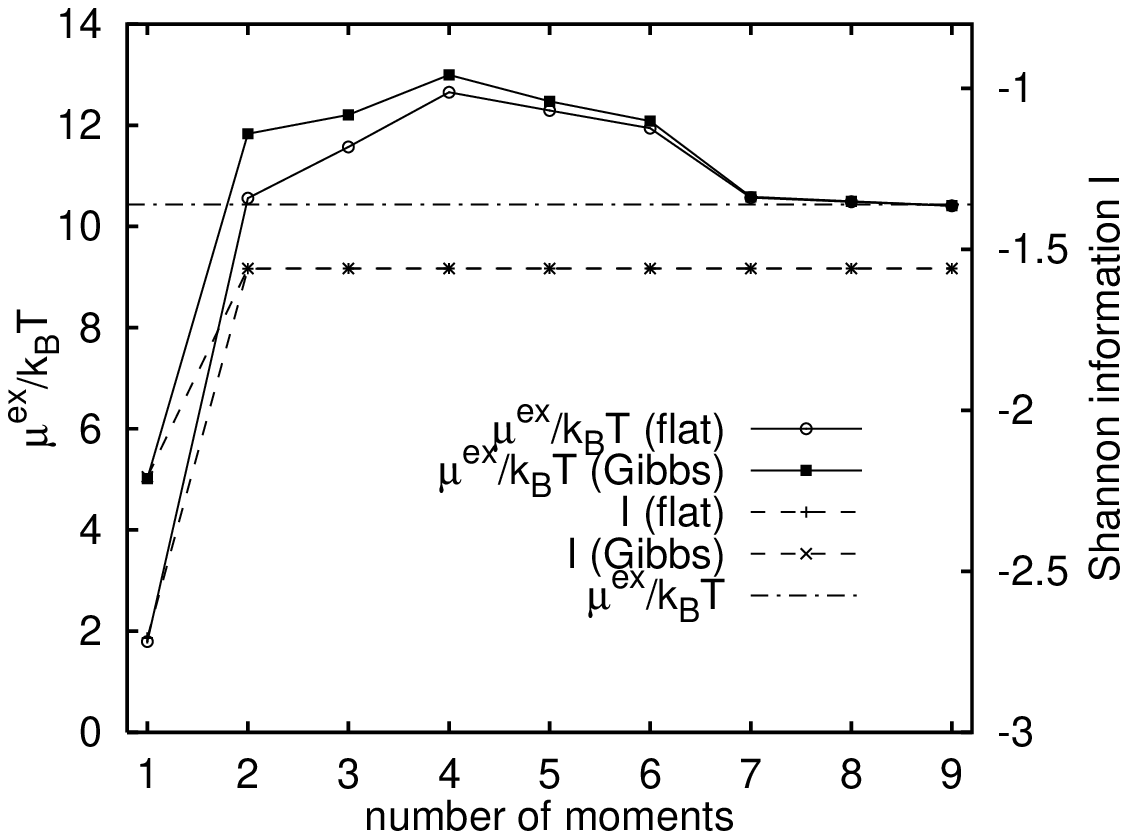}}
  \caption{Effect of increasing the number of moments used in the
    IT calculations.  Shown as a function of the
    number of moments is the predicted excess chemical potential of a
    methane-size hard sphere with exclusion radius $d=0.33$~nm for
    solvation in SPC/E water\protect\cite{Berendsen:87} (left-hand
    scale).  Also shown for reference is the actual chemical potential
    (dot-dashed line).  The Shannon information $I$ (right-hand scale)
    illustrates that the third and higher moments do not result in an
    appreciable gain in information.}
  \label{fig:moments}
\end{figure}
}

\def\FIGfour{
\begin{figure}[htbp]
\centerline{\psboxto(6cm;0pt){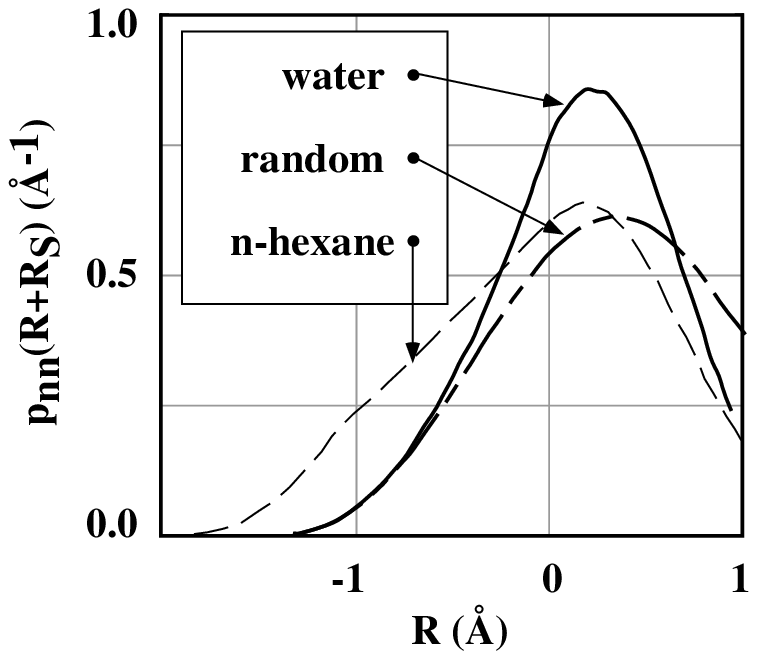}}
  \caption{Distributions $p_{nn}$ of the distance $R+R_S$ from a random point
    to the nearest neighboring site in water, n-hexane, and an ideal
    gas with water density.\protect\cite{Palma} $R_S$ is the van der
    Waals radius of the solvent.  For each of the three fluids, the
    solvent van der Waals volumes are represented by a superposition
    of spherical sites of one type only.  The ideal-gas curve is the
    Hertz distribution $4\pi\rho(R+R_S)^2 \exp[-4\pi\rho(R+R_S)^3 /3]$
    with $\rho$ the molecular (or oxygen) density of liquid water at
    301~K and atmospheric pressure.  The van der Waals radius of water
    oxygen is $R_S=1.35$ {\AA}; that of united-atom carbons in
    n-hexane is $R_S=1.85$ {\AA}.  }
  \label{fig:bklee}
\end{figure}
}

\def\FIGfive{
\begin{figure}[htbp]
\centerline{\psboxto(7cm;0pt){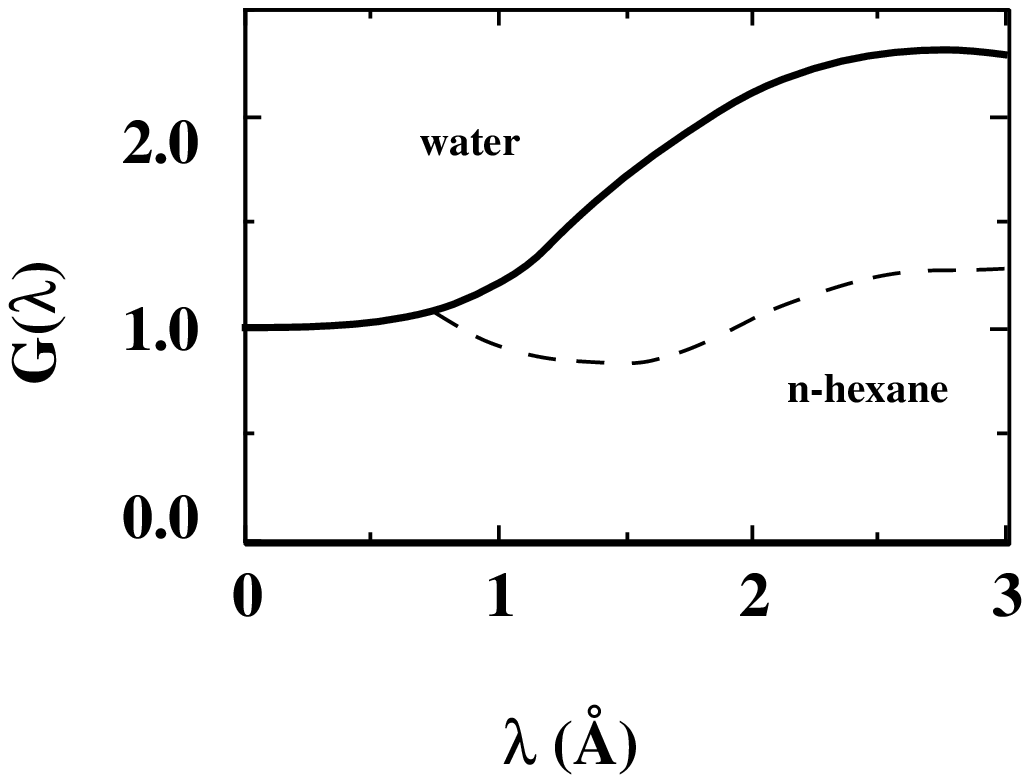}}
  \caption{Contact densities $G(\lambda)$ of water and n-hexane at the
    surface of hard-sphere solutes.\protect\cite{Pratt:PNAS:92}
    $G(\lambda)$ gives the compressive force of the solvent on a hard
    spherical solute.}
\label{fig:Gforce}
\end{figure}
}

\def\FIGsix{
\begin{figure}[htbp]
\centerline{\psboxto(7cm;0pt){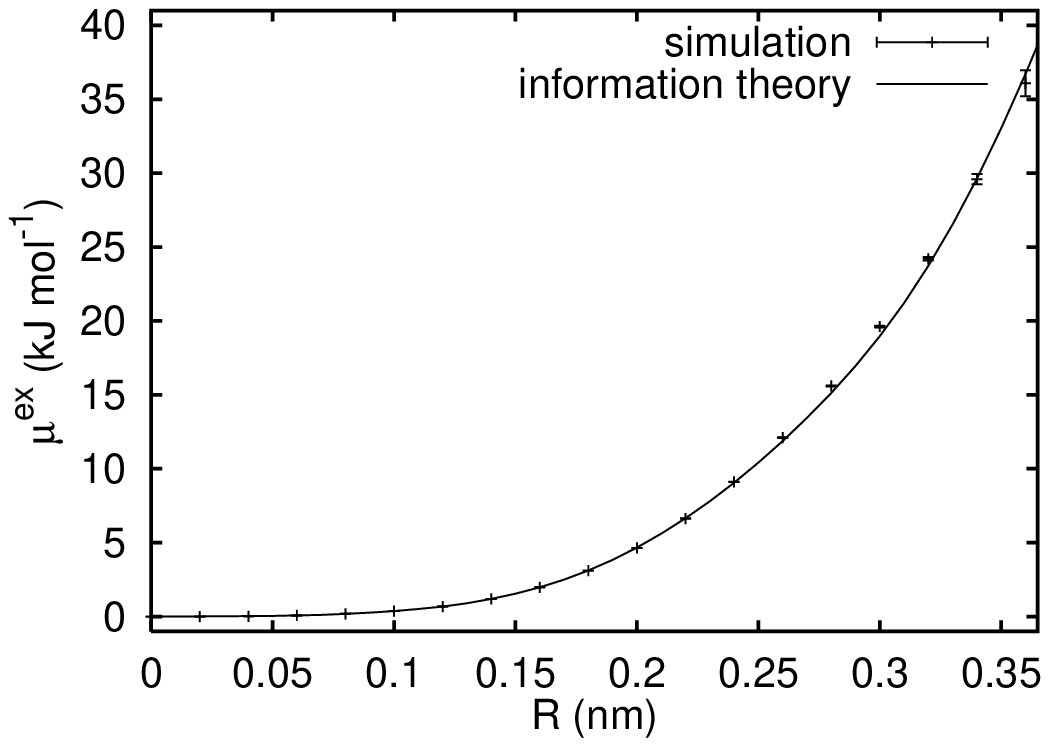}}
  \caption{Excess chemical potential of hard-sphere solutes in SPC
    water\protect\cite{Berendsen:81} as a function of the exclusion
    radius $d$.  The symbols are simulation results, compared with the
    IT prediction using the flat default model (solid
    line).\protect\cite{Hummer:PNAS:96}}
  \label{fig:sphere_mu}
\end{figure}
}

\def\FIGseven{
\begin{figure}[htbp]
\centerline{\psboxto(7cm;0pt){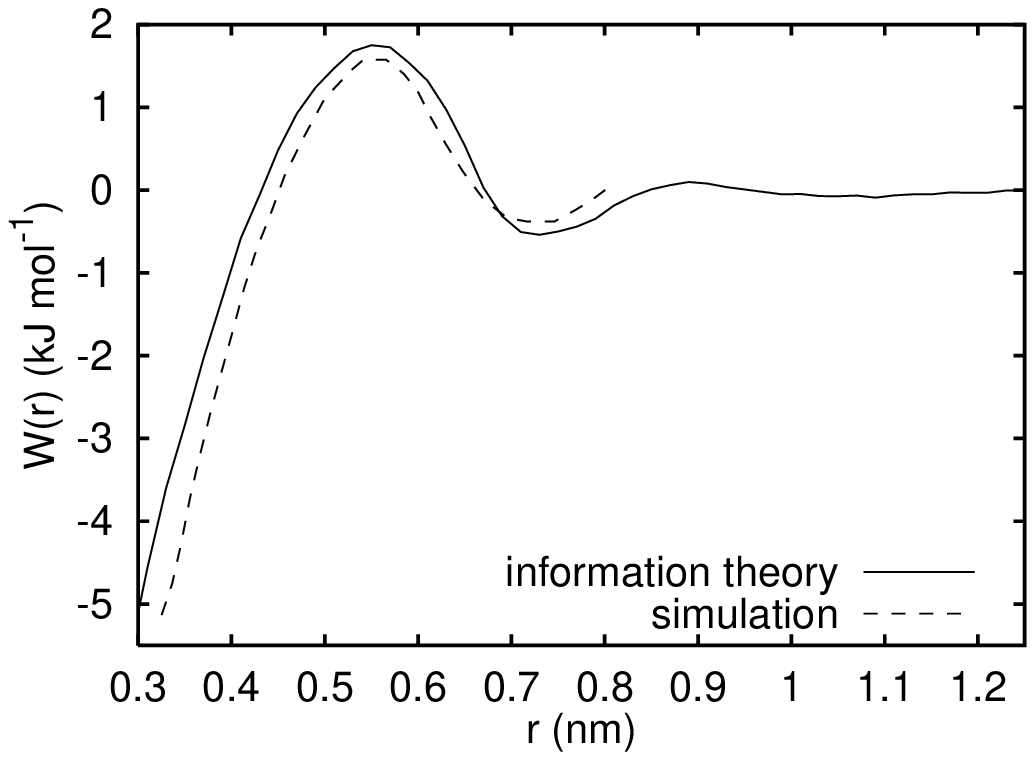}}
  \caption{PMF between two methane-size cavities
    in SPC water,\protect\cite{Berendsen:81} comparing explicit
    computer simulations\protect\cite{Smith:JCP:93} (dashed line) with
    the IT prediction\protect\cite{Hummer:PNAS:96} (solid line).  Note
    that the simulation curve is based on continuous solute-solvent
    interactions, subtracting the methane-methane interaction
    potential from the methane-methane PMF.}
  \label{fig:PMF}
\end{figure}
}

\def\FIGeight{
\begin{figure}
\centerline{\psboxto(7cm;0pt){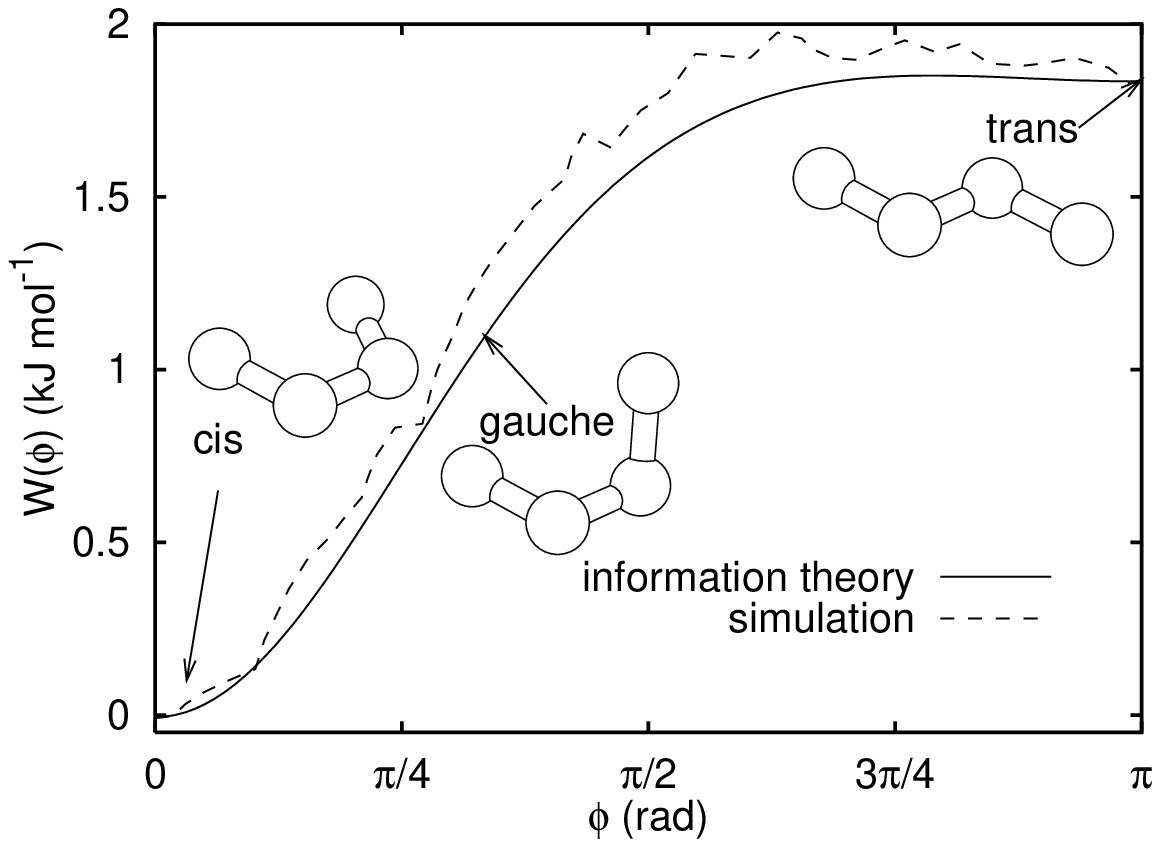}}
  \caption{Water contribution to the torsional PMF
    of butane from IT\protect\cite{Hummer:PNAS:96} and
    explicit computer simulation\protect\cite{Beglov:94} (dashed
    line).}
  \label{fig:butane}
\end{figure}
}

\def\FIGnine{
\begin{figure}[htbp]
\centerline{\psboxto(7cm;0pt){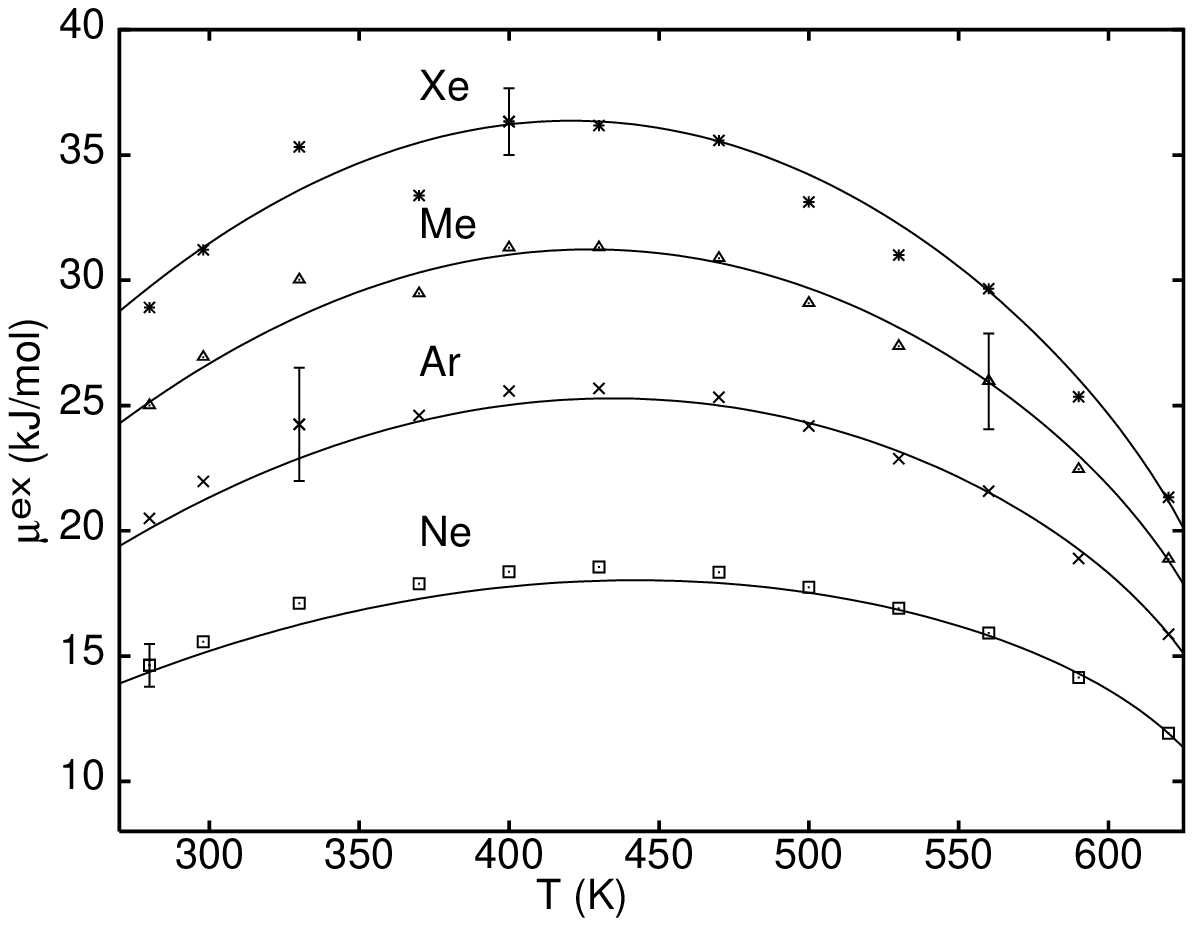}}
    \caption{Chemical potential of hard-sphere solutes with exclusion
      radii corresponding to neon, argon, methane, and xenon as a
      function of temperature along the saturation curve of
      water.\protect\cite{Garde:PRL:96} The solid line is the IT
      prediction.  The symbols are computer simulation data from
      insertion.  Both simulation and theory data are based on SPC
      water\protect\cite{Berendsen:81} data.}
  \label{fig:muT}
\end{figure}
}

\def\FIGten{
\begin{figure}
\centerline{\psboxto(7cm;0pt){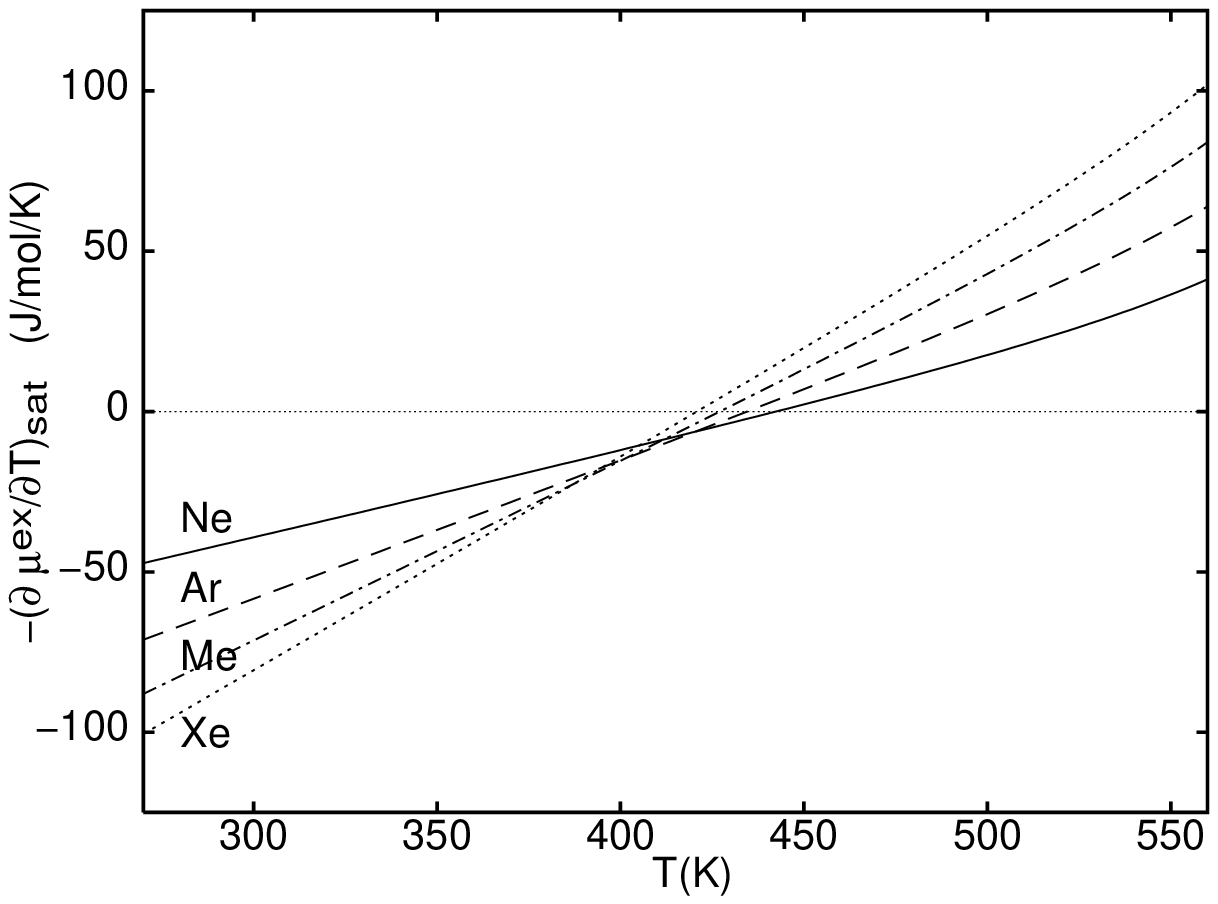}}
\caption{Hydration entropy of hard-sphere solutes with exclusion
  radii corresponding to neon, argon, methane, and krypton as a
  function of temperature along the saturation curve of
  water.\protect\cite{Garde:PRL:96}  Shown is the entropy defined in
  eq~\protect\ref{eq:ent}.}
\label{fig:ent}
\end{figure}
}

\def\FIGeleven{
\begin{figure}[htbp]
\centerline{\psboxto(7cm;0pt){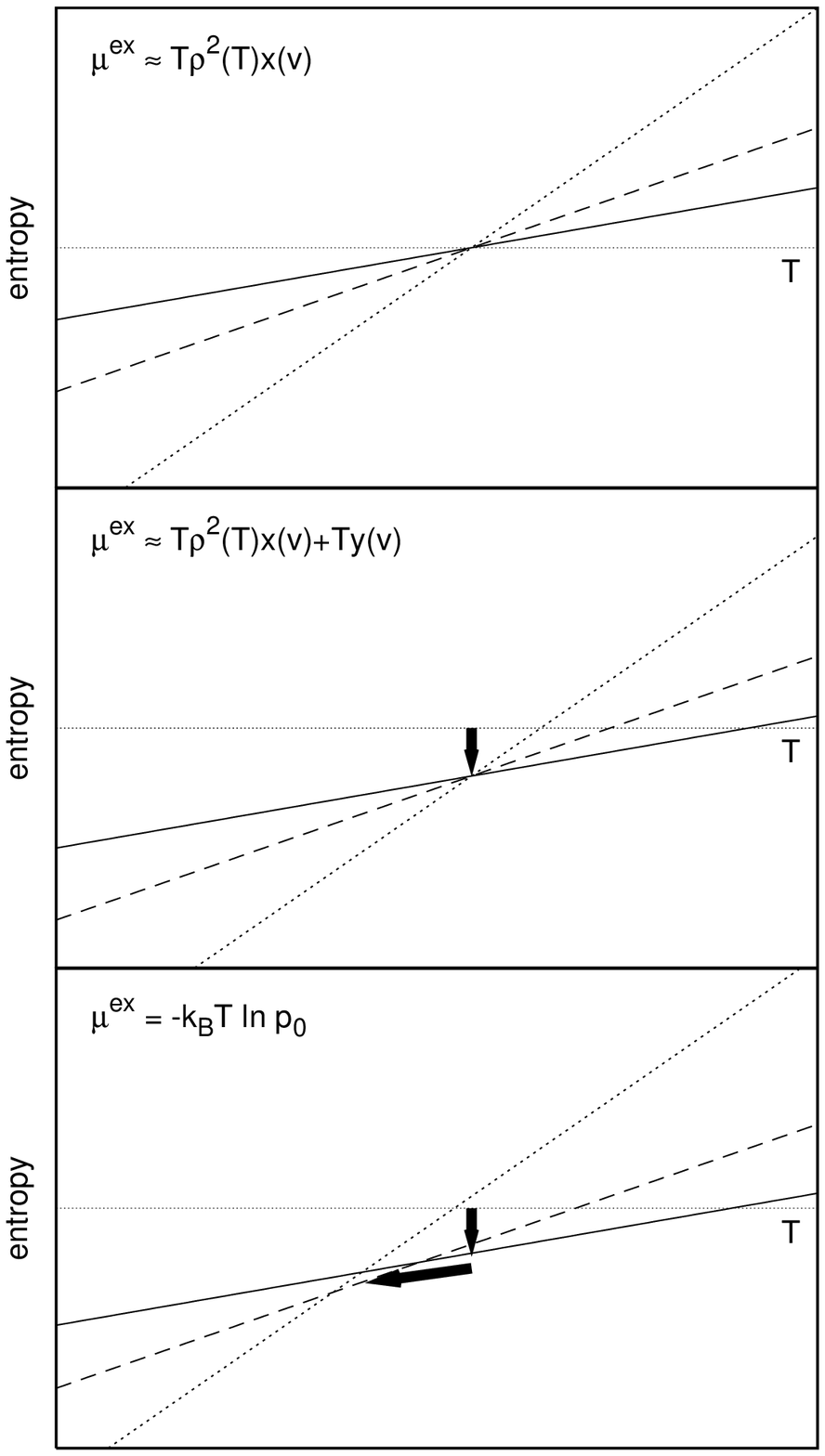}}
  \caption{Schematic representation of the different factors resulting
  in approximate convergence of solvation entropies.}
  \label{fig:ent_schem}
\end{figure}
}

\def\FIGtwelve{
\begin{figure}[htbp]
\centerline{\psboxto(8cm;0pt){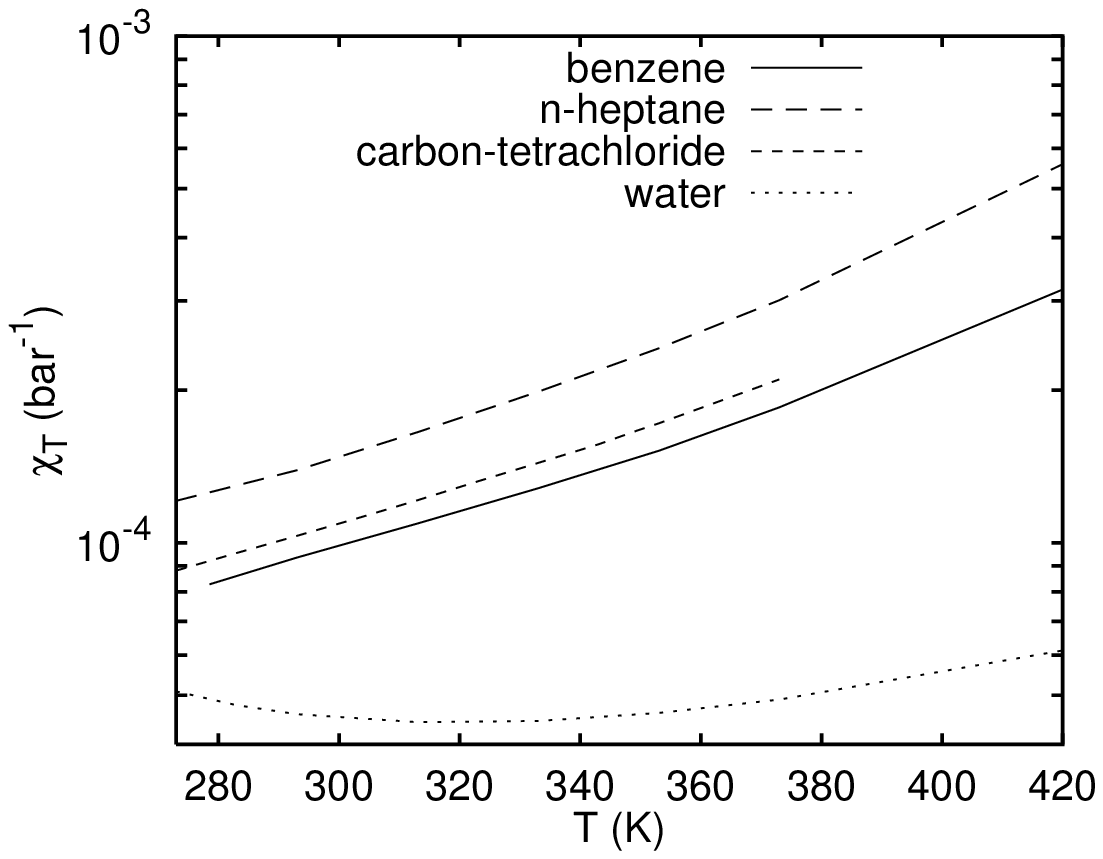}}
  \caption{Comparison of the isothermal compressibility of water and
    nonpolar solvents as a function of temperature along their
    respective saturation curves.  Data are taken from
    Ref.~\protect\onlinecite{Rowlinson:Swinton:82}.}
  \label{fig:comp}
\end{figure}
}

\def\FIGthirteen{
\begin{figure}[htbp]
\centerline{\psboxto(7cm;0pt){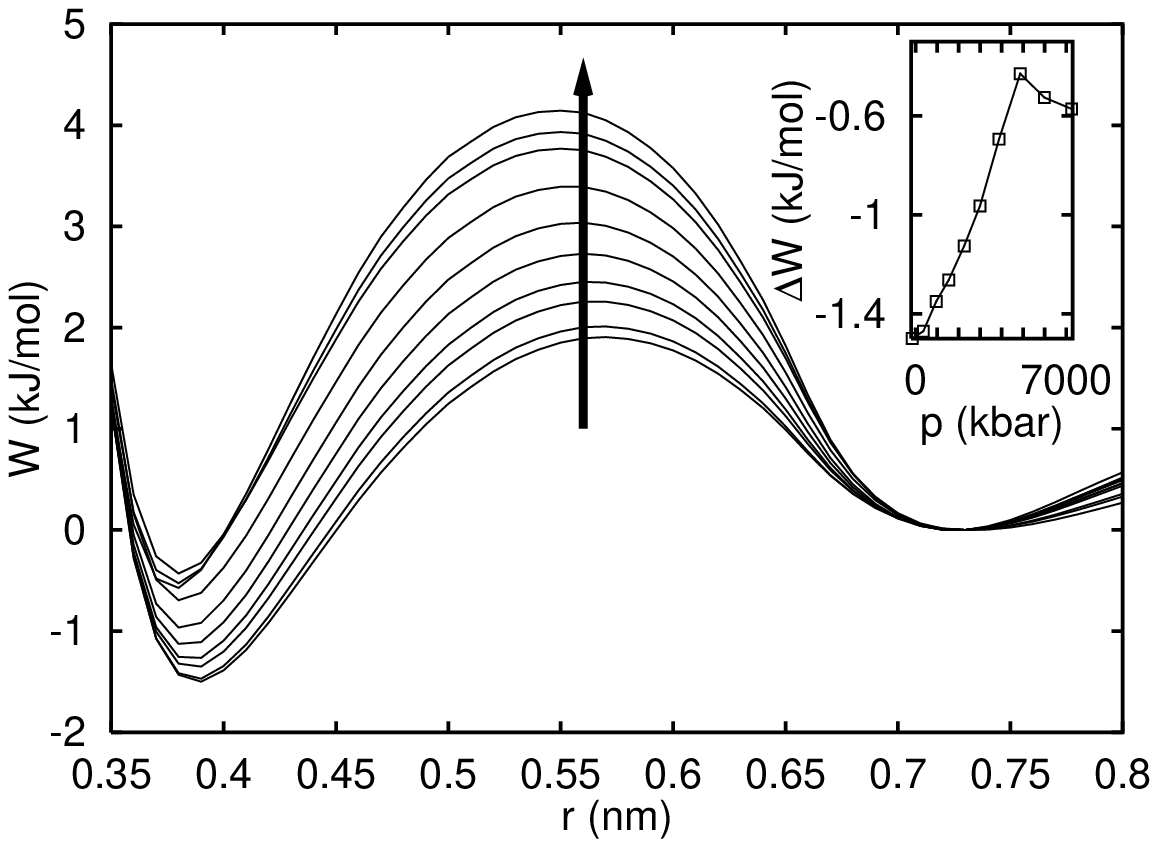}}
  \caption{Pressure dependence of the PMF between two methane-like
    solutes in water.  The PMF's were calculated from information
    theory using a flat default model.\protect\cite{Hummer:PNAS:98} A
    methane-methane Lennard-Jones interaction was added.  SPC
    water\protect\cite{Berendsen:81} $g(r)$'s were used.  The arrows
    indicate changes with increasing pressures from $-160$ to 7250
    bar.}
  \label{fig:Wp}
\end{figure}
}

\def\FIGfourteen{
\begin{figure}[htbp]
\centerline{\psboxto(7cm;0pt){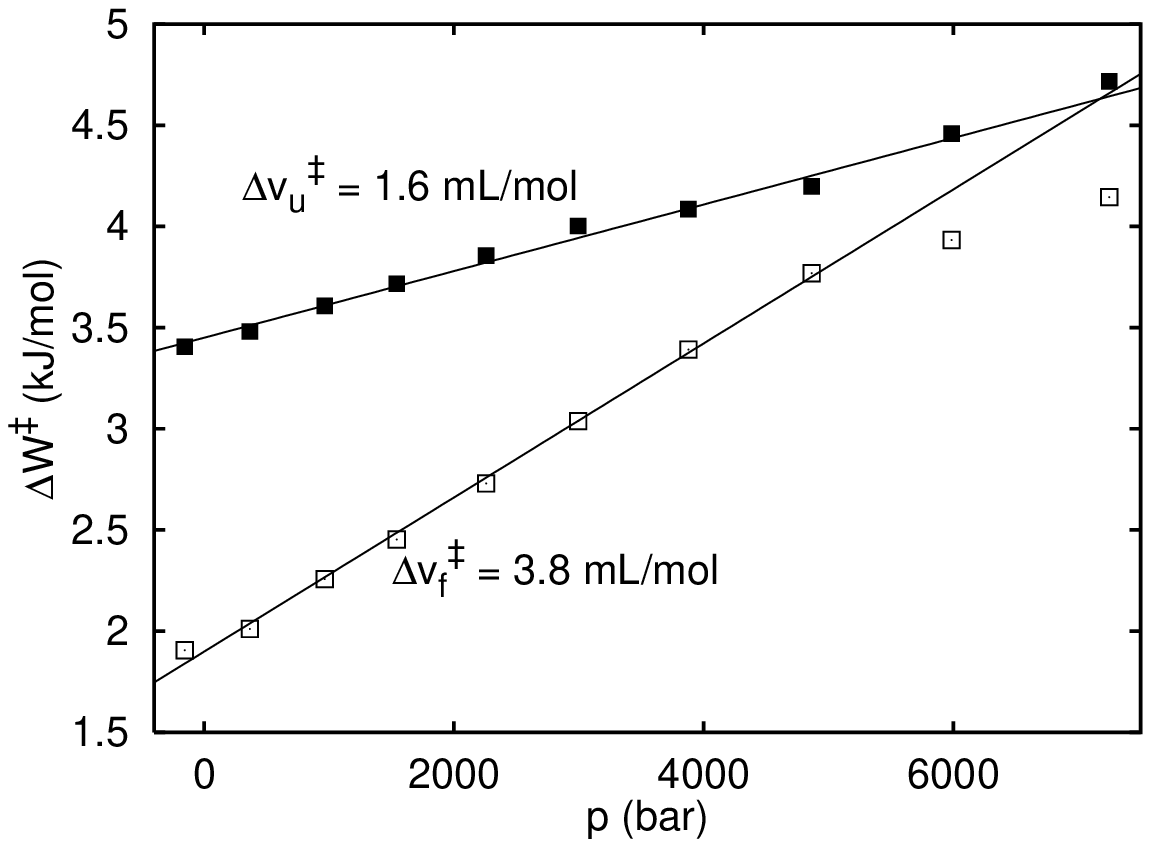}}
  \caption{Pressure dependence of the desolvation barrier between the
    contact and solvent-separated minimum in the methane-methane PMF
    of Figure~\protect\ref{fig:Wp}.  The activation free energy for
    forming (open squares) and breaking hydrophobic contacts (filled
    squares) is shown as a function of pressure.}
  \label{fig:Wbarr}
\end{figure}
}

\def\FIGfifteen{
\begin{figure}[htbp]
\centerline{\psboxto(7cm;0pt){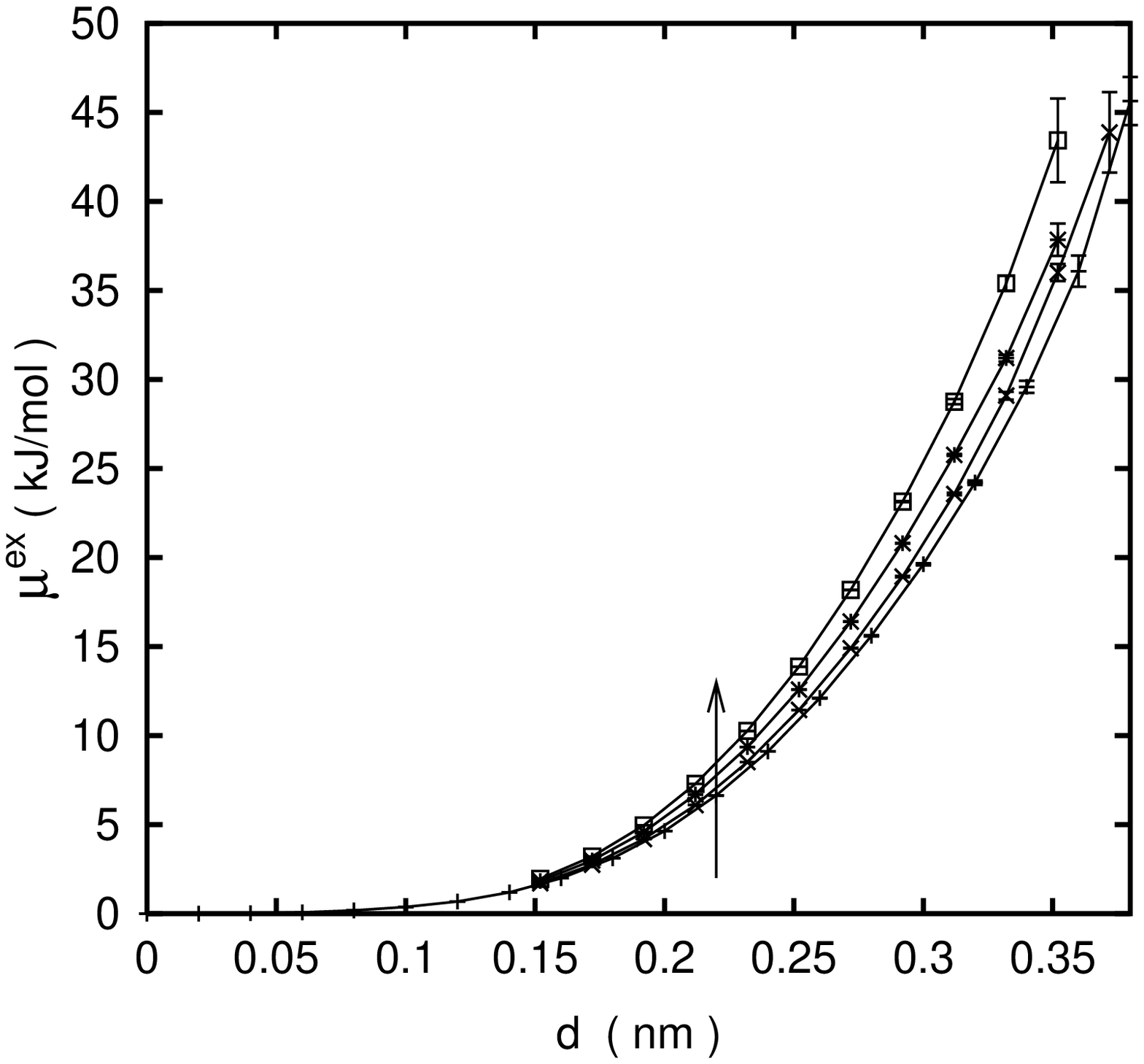}}
  \caption{Excess chemical potential of hard-sphere solutes in
    aqueous NaCl solution as a function of solute size, given by the
    exclusion radius with water, and salt concentration.}
  \label{fig:salt}
\end{figure}
}

\begin{document}

\title{Hydrophobic Effects on a Molecular Scale}

\author{G. Hummer,$^*$ S. Garde, A. E. Garc\'{\i}a, M. E.
  Paulaitis,$^\dagger$ and L. R. Pratt}

\address{Theoretical Division, Los Alamos National Laboratory, Los
  Alamos, New Mexico 87545, USA\\
  $^\dagger$Department of Chemical Engineering, Johns Hopkins University,
  Baltimore, Maryland 21218, USA\\[.2\baselineskip]
  $^*$ Corresponding Author.  Mail Stop K710. Phone: (505) 665-1923.
  Fax: (505) 665-3493. E-mail: hummer@lanl.gov}

\date{LA-UR 98-2758}

\def\Abstract{
  A theoretical approach is developed to quantify hydrophobic
  hydration and interactions on a molecular scale, with the goal of
  insight into the molecular origins of hydrophobic effects.  The
  model is based on the fundamental relation between the probability
  for cavity formation in bulk water resulting from molecular-scale
  density fluctuations, and the hydration free energy of the simplest
  hydrophobic solutes, hard particles.  This probability is estimated
  using an information theory (IT) approach, incorporating experimentally
  available properties of bulk water -- the density and radial
  distribution function.  The IT approach reproduces
  the simplest hydrophobic effects: hydration of spherical nonpolar
  solutes, the potential of mean force (PMF) between methane molecules, and
  solvent contributions to the torsional equilibrium of butane.
  Applications of this approach to study temperature and pressure
  effects provide new insights into the thermodynamics and kinetics of
  protein folding.  The IT model relates the
  hydrophobic-entropy convergence observed in protein 
  unfolding experiments to the macroscopic isothermal compressibility
  of water.
  A novel explanation for pressure
  denaturation of proteins follows from an analysis of the pressure
  stability of hydrophobic aggregates, suggesting that water
  penetrates the hydrophobic core of proteins at high pressures.  This
  resolves a long-standing puzzle, whether pressure denaturation
  contradicts the hydrophobic-core model of protein
  stability.  Finally, issues of ``dewetting'' of molecularly large
  nonpolar solutes are discussed in the context of a recently
  developed perturbation theory approach.
}

\maketitle

\section{Introduction}
Hydrophobic interactions play a central role in many self-assembly
processes in aqueous solution: protein folding; the formation of
micelles, membranes, and complex mesophases in surfactant
solutions; the formation of macromolecular complexes; and the
non-specific aggregation of proteins in inclusion bodies during
over-expression or in disease-causing amyloid plaque formation in
brain tissue.
\cite{Kauzmann:APC:59,Tanford:80,Dill:B:90,Blokzijl:AC:93,Fink:FD:98}
Biopolymers such as proteins, nucleic acids, and lipids contain a
significant fraction of nonpolar groups: aliphatic and aromatic
amino-acid side chains, the faces of nucleic-acid bases, and the
hydrocarbon tails of lipids.  Attractive hydrophobic interactions
between these nonpolar groups contribute significantly to the
stability of folded proteins, base-stacking in helical nucleic acids,
and membrane bilayers.  Protein folding may be the most extensively
characterized macromolecular self-assembly process in aqueous
solution.
Thermodynamic\cite{Kauzmann:APC:59,Tanford:80,%
Privalov:APC:79,Privalov:APC:88,Makhatadze:APC:95} and structural
studies\cite{Richards:JMB:74,Buckle:JMB:93} on proteins have shown
that the formation of a hydrophobic core, comprising predominantly
nonpolar amino acids, plays a significant if not the dominating role in 
protein folding.  Clearly, other interactions, such as electrostatic 
interactions and intramolecular
or water-mediated hydrogen bonding, cannot be neglected in a full
treatment of the thermodynamic stability of amphiphilic molecules and their
assemblies.  Nevertheless, a quantitative understanding of the 
stability of hydrated amphiphilic macromolecules necessitates a 
thorough, quantitative description of hydrophobic driving forces as a major
contributing factor.

Despite many decades of research on hydrophobic effects, our
understanding of key phenomena presumably of hydrophobic origin is
still incomplete, and seemingly contradictory observations have yet to
be explained.  Surprisingly, these include the effects of changing
even the most elementary thermodynamic variables, temperature and
pressure, or key solution properties, such as salt concentration and
composition.  The well-known fact that proteins
denature at elevated temperatures and at elevated pressures is an
immediate example.  Heat
denaturation can be explained by the hydrophobic core model of
proteins, which was established on the basis of extensive temperature
unfolding studies.\cite{Privalov:APC:79,Privalov:APC:88,%
  Makhatadze:APC:95,Baldwin:PNAS:86,Baldwin:PNAS:92} These studies
show that the thermodynamic characteristics of heat denaturation are
similar to those of transferring nonpolar molecules from a
hydrocarbon or gas phase into water.  This suggests that nonpolar
residues in the protein interior become exposed to solvent during
denaturation.  However, as pointed out poignantly by Kauzmann,
\cite{Kauzmann:87} pressure denaturation appears to contradict the
hydrophobic-core model of proteins in the sense that the hydrocarbon
transfer process and protein denaturation exhibit diametrically opposite
pressure dependences of the associated volume change.

One goal of our work has been to reconcile these apparently
contradictory results for temperature and pressure denaturation of
proteins within the context of the hydrophobic core model of protein
folding.  To achieve this, we developed detailed molecular
explanations for the underlying
processes.\cite{Garde:PRL:96,Hummer:PNAS:98}  Here, we will focus specifically
on the observed temperature convergence of hydrocarbon-transfer
entropies from calorimetry experiments.\cite{Garde:PRL:96} That
entropy convergence coincides with the observed temperature
convergence of the entropy of protein unfolding and thus provides a
thermodynamic foundation for the hydrophobic-core model of
protein folding.\cite{Privalov:APC:79,Privalov:APC:88,Baldwin:PNAS:86,%
  Baldwin:PNAS:92,Privalov:JMB:74,Makhatadze:95,Privalov:JMB:93} In
addition, we will suggest a molecular mechanism for pressure
denaturation of
proteins\cite{Brandts:70,Zipp:73,Heremans:82,Weber:83,Silva:93,%
  Jonas:94,Royer:95,Silva:97} and discuss its consequences on folding
kinetics and the characteristics of the ensemble of unfolded protein
structures.\cite{Hummer:PNAS:98} Our explanation of pressure
denaturation invokes experimental observations regarding the
differences in the structures of heat and pressure-denatured proteins,
the latter being more compact.\cite{Silva:93,Zhang:95,Panick:98} We
will also give preliminary results for the effects of salt
concentration on hydrophobic hydration, as characterized by the
Hofmeister series,\cite{Baldwin:BJ:96} which ranks salts by their ability
to increase (``salt in'') or decrease (``salt out'') the solubility of
nonpolar molecules in water.

To achieve these goals, we need a model that quantifies
hydrophobic hydration and hydrophobic
interactions, and the effects of temperature, pressure, and salt
concentration on them.  A number of models of the hydrophobic effect
have been developed in the
past,\cite{Kauzmann:APC:59,Tanford:80,Blokzijl:AC:93,Frank:JCP:45,%
  Nemethy:JCP:62,Ben-Naim:80,Besseling:JPCB:97,Madan:JPCB:97} all
having their merits and limitations.\cite{LitNote} One of the most
influential models is scaled-particle theory (SPT)\cite{Reiss:59} and
its extension to hydrophobic
hydration.\cite{Pierotti:63,Stillinger:73} SPT laid a molecular
foundation for surface area models of hydrophobic hydration with
macroscopic surface
tensions.\cite{Hermann:JPC:72,Chothia:N:74,Oobatake:PBMB:93,Sharp:S:91,Simonson:94,Giesen:94,Tunon:94,Fukunishi:JPC:96}
However,
this invocation of macroscopic parameters limits applications
of SPT to study the subtle changes due to temperature and pressure,
for example, that are closely associated with the molecule-scale
properties.  A thoroughly molecular theory was developed by Pratt and
Chandler (PC),\cite{Pratt:JCP:77} based on the theory of simple
liquids.  The PC theory invokes radial correlation functions in the
context of Ornstein-Zernike integral equations with appropriate
closures.  Lazaridis and Paulaitis\cite{Lazaridis:92} (LP) use an
expansion of the conformational entropy in terms of particle
distribution functions.  Both PC and LP theories relate the structural
ordering of water by the nonpolar solute to the hydration
thermodynamics through solute-water pair correlations.

Probably the most powerful approach is the direct computer simulation
of hydrophobic hydration phenomena using molecular dynamics or Monte
Carlo methods.\cite{Pangali:79} Computer simulations are flexible and,
with the advent of ever faster computers, the system size and time
scale limitations are becoming less of a concern.  However, the {\em
  interpretation} of computer simulation results necessitates a
well-founded theoretical framework.  Here, and in the aforementioned
theories, computer simulations provide critical input that is not
readily available from experiments.

The theoretical approach pursued here describes hydrophobic hydration
and hydrophobic interactions using chemical potentials of the simplest
hydrophobic solutes in water -- ``hard'' molecules exerting entirely
repulsive interactions on water molecules.  The solvation chemical potentials
for hard-core solutes are directly related to the presence of cavity
volumes of molecular size due to density fluctuations in bulk water.
We quantify these density fluctuations using an approach motivated by
information theory (IT).\cite{Hummer:PNAS:96}
From a maximum-entropy principle,\cite{Jaynes:83}
we find the ``best-possible'' description of the
fluctuations that satisfies certain experimental constraints.  The
virtues of such an approach are its simplicity, efficiency, and
accuracy for molecule-size solutes.  In addition, the IT
model expresses the simple hydrophobic phenomena using
properties of bulk water alone.  This allows us to relate hydrophobic
effects to the peculiar properties of water that distinguishes it
from nonpolar solvents.  The IT model builds on
concepts introduced in SPT, specifically the relation between the
probability of finding a molecule-size cavity in water and the
chemical potential of solvation of hard particles.  In its simplest
form, the IT model is also related to Ornstein-Zernike integral
equation theories and the PC
theory.\cite{Hummer:PNAS:96,Chandler:PRE:93,Berne:PNAS:96,Percus:JP:93}

The paper is organized as follows: After developing the IT
model for hard nonpolar solutes and a generalization to
continuous solute-solvent interactions, we will discuss practical
aspects of the implementation.  Limitations in the macroscopic limit
will be analyzed before we view the IT model from a
historical perspective.  We will then present results for the simplest
hydrophobic effects -- solvation of spherical solutes, methane-methane
interaction in water, and contributions to the torsional equilibrium
of butane.  The temperature dependence of hydrophobic hydration and
the entropy convergence will be studied subsequently.  We will then
quantify the effects of pressure on hydrophobic interactions, leading
to a model for the pressure denaturation of proteins.  The effects of
salt on hydrophobic hydration will be discussed briefly.  We will also
introduce a recently developed perturbation theory
model\cite{Hummer:PRL:98} based on the energetic loss of water
molecules at the solute-water interface.  This model allows us to
extend our analysis of hydrophobic effects further to mesoscopic and
macroscopic solutes.  We will conclude with an attempt to give answers
to the question how water differs from hydrocarbon liquids as a solvent
for nonpolar solutes.

\section{Information theory model of hydrophobic hydration}
\subsection{Simple hydrophobic effects: thermodynamics of
  dissolving a hard particle in water} The simplest hydrophobic solute
is a cavity or, equivalently, a hard-core particle which excludes
water-oxygen atoms from a volume $v$ of a given shape and molecular size.
Cavity formation constitutes the important, first step in solvating any
nonpolar solute in water.  The solvation thermodynamics of a hard
particle is determined by the excess chemical potential $\mu^{\rm ex}$
corresponding to the free energy of its transfer from an ideal gas into
the aqueous environment.  Statistical mechanics relates this excess
chemical potential to the probability $p_0$ of finding an empty volume $v$
in water, i.e., a cavity of a given size and shape,\cite{Reiss:59,Pollack:S:91}
\begin{equation}
  \mu^{\rm ex} = -k_{\rm B}T \ln p_0~.
  \label{eqn:ktlnp}
\end{equation}
To determine the chemical potential $\mu^{\rm ex}$, we need to
quantify the probability $p_0$ of successfully inserting a hard-core
solute of a given size and shape into equilibrium conformations of
water, as illustrated in
Figure~\ref{fig:pzero}\nocite{Hummer:PNAS:96,Berendsen:81}.  A virtue
of such an approach is that the solvation thermodynamics characterized
by $\mu^{\rm ex}$ is determined by the properties of pure water, with
the solute entering through its molecular volume $v$.

\if T\TS \FIGone \fi

\subsection{Information theory approach to excess chemical
  potentials of solvation} The goal is to estimate accurately the
probability $p_0$ that a given hard-core solute inserted into water
does not overlap with any of the solvent centers, defined as the
positions of water-oxygen atoms.  The IT approach does
not attempt to model this quantity directly. Instead, it focuses on
the set of probabilities $p_n$ of finding $n$ water-oxygen atoms
inside the observation volume, with $p_0$ being just one of the $p_n$.
We will attempt to get accurate estimates of the $p_n$, and $p_0$ in
particular, using experimentally available information as constraints
on the $p_n$.  The moments of the fluctuations in the number of
solvent centers within the observation volume $v$ provide such
constraints.

For a given observation volume $v$ in bulk water, the moments of the
fluctuations in the particle number $n$ are determined from the $p_n$
as follows:
\begin{eqnarray}
  \label{eq:fluc}
  \langle n^k\rangle & = & \sum_{n=0}^\infty p_n n^k~,
\end{eqnarray}
where $\langle\cdots\rangle$ denotes a thermal average. The zeroth,
first, and second moment can be expressed in terms of
experimentally accessible quantities:
\begin{mathletters}
  \label{eq:moments}
  \begin{eqnarray}
    \label{eq:momzero}
    \langle 1\rangle & = & \sum_{n=0}^\infty p_n = 1\\
    \label{eq:momone}
    \langle n \rangle & = &\sum_{n=0}^\infty p_n n = \rho v~,\\
    \label{eq:momtwo}
    \langle n^2\rangle & = & \sum_{n=0}^\infty p_n n^2 = \langle
    n\rangle + \rho^2 \int_{v}d{\rm{\bf
        r}} \int_{v}d{\rm{\bf r}'}g(|{\rm{\bf r} -{\bf r}'}|)~,
  \end{eqnarray}
\end{mathletters}
where $\rho$ is the number density of bulk water and $g(r)$ is the
radial distribution function between water-oxygen atoms in bulk water,
which can be determined from X-ray or neutron scattering measurements or
computer simulations.  These moment conditions provide
constraints on the $p_n$'s, and guarantee that the $p_n$ are normalized and
have the correct first and second moments.  Higher moments would
require knowledge of triplet and higher-order correlation
functions which are not generally accessible to experiments, but can
be calculated from computer simulations.\cite{Hummer:ice:94}

The IT approach attempts to provide the
``best-possible'' estimate of the probabilities $p_n$ under the
constraints of the available information,\cite{Jaynes:83}
defined as the set $\{p_n\}$
that maximizes an information entropy $\eta$ subject to the
information constraints,
\begin{eqnarray}
  \label{eq:maxent}
  \max_{\rm \{constraints\}} \eta(\{p_n\})~.
\end{eqnarray}
In the most general form, we adopt a relative or cross
entropy,\cite{Shore:80}
\begin{eqnarray}
  \eta(\{p_n\}) & = & - \sum_{n=0}^{\infty} p_n \ln
  \left(\frac{p_n}{\hat{p}_n}\right)~,
  \label{eq:entropy}
\end{eqnarray}
where $\hat{p}_n$ represents an empirically chosen ``default model.''

We consider two natural choices of default models: the Gibbs default
model $\hat{p}_n\propto 1/n!$, which leads to a Poisson distribution for a
given mean, as would be expected for an ideal gas, and a flat
distribution ($\hat{p}_n=1$ for $n\leq n_{\rm max}$ and $\hat{p}_n=0$
otherwise), which results in a discrete Gaussian form of $p_n$ with given
mean and variance.  Empirically, we find that the latter choice is accurately
applicable to molecule-size cavities.\cite{Hummer:PNAS:96}

\if T\TS \FIGtwo \fi

Maximizing the information entropy under the constraints of
eq~\ref{eq:moments} leads to
\begin{eqnarray}
  \label{eq:pnmax}
  p_n & = & \hat{p}_n e^{\lambda_0+\lambda_1 n+\lambda_2 n^2}~,
\end{eqnarray}
where $\lambda_0$, $\lambda_1$, and $\lambda_2$ are the Lagrange
multipliers chosen to satisfy the moment conditions eq~\ref{eq:moments}.

Figure~\ref{fig:sphere_pn} shows $p_n$ distributions for spherical
observation volumes $v$ calculated from computer simulations of
SPC\cite{Berendsen:81} water.\cite{Hummer:PNAS:96} The solute
exclusion volume is defined by the distance $d$ of closest approach of
water-oxygen atoms to the center of the sphere.  For the range of
solute sizes studied, we find that $\ln p_n$ values are closely
parabolic in $n$.  This would be predicted from the flat default model
with $n_{\rm max}\rightarrow\infty$, as shown in
Figure~\ref{fig:sphere_pn}.

\if T\TS \FIGthree \fi

Figure~\ref{fig:moments}\nocite{Berendsen:87} illustrates the effect
of including higher moments in the IT model.  Results
are shown both for the Gibbs and the flat
default models.  We find that the prediction of
$\mu^{\rm ex}$ is greatly improved when the second moment (i.e., the
variance of the particle number) is included in addition to the mean.
However, inclusion of higher moments initially makes the prediction
worse. Only when seven or more moments are used is the prediction as
accurate as the two-moment model.  Also shown in
Figure~\ref{fig:moments} is the calculated Shannon information $I(\{p_n\})$,
\begin{eqnarray}
  \label{eq:Shannon}
  I(\{p_n\}) & = & \sum_{n=0}^\infty p_n \ln p_n~.
\end{eqnarray}
Including the second moment results in a large gain of information,
whereas the gains from including additional higher moments are small:
moments of order three and higher are, in this respect,
``un-informative,'' as measured by the gain in Shannon information
$I(\{p_n\})$ (see also Ref.~\onlinecite{Pratt:NATO:98}).  Those higher
moments do not significantly alter the estimated values of $p_0$ and
$\mu^{\rm ex}$ for the solute-size range considered in
Figures~\ref{fig:sphere_pn} and \ref{fig:moments}.

\subsection{Continuous information theory}
\label{sec:cont} 
A more general statement of the principles underlying the IT
approach follows from a generalization to continuous
interactions between the solvent and the hydrophobic solute.  Such a
generalization can be achieved by explicitly considering solvent
positions within a microscopic volume, in addition to the occupancy
numbers considered in the basic IT approach.  We
define probability distributions $p(j;{\bf r}_1,\ldots,{\bf r}_j)$ of
observing exactly $j$ particles within the observation volume $v$ at
positions ${\bf r}_1,\ldots,{\bf r}_j$ in infinitesimal volume elements $d{\bf
  r}_1,\ldots,d{\bf r}_j$.  The $p(j;{\bf r}_1,\ldots,{\bf r}_j)$
allow us to calculate solvation chemical potentials of a solute with
continuous interactions $u({\bf r}_1,{\bf r}_2, \ldots)$ between
solvent molecules at positions ${\bf r}_1,{\bf r}_2,\ldots$ and a
solute at the origin.  Within the observation volume $v$, the
solute-solvent interactions are treated explicitly.  Finite-range
interactions provide a natural choice for $v$; otherwise, a finite
volume $v$ can be chosen with corrections for long-range
interactions.

Widom's formula\cite{Widom:63} relates the excess chemical potential
of the solute $\mu^{\rm ex}$ to the normalized $p(j;{\bf
  r}_1,\ldots,{\bf r}_j)$,
\begin{eqnarray}
  \label{eq:Widom}
  \lefteqn{
  e^{-\beta\mu^{\rm ex}}  = \left\langle e^{-\beta u({\bf r}_1,{\bf
  r}_2,\ldots)}\right\rangle}\nonumber\\
  & = & \sum_j \left[ \prod_{i=1}^{i=j} \int_v
    d{\bf r}_i \right] p(j;{\bf r}_1,\ldots,{\bf r}_j)e^{-\beta u({\bf
        r}_1,\ldots,{\bf r}_j)}~,
\end{eqnarray}
where $\beta^{-1} = k_{\rm B}T$.

Calculating $p(j;{\bf r}_1,\ldots,{\bf r}_j)$ directly would require
knowledge of higher-order correlation functions.  Instead, our goal is
to infer the $p(j;{\bf r}_1,\ldots,{\bf r}_j)$ from available
information, such as the one- and two-particle densities in the
observation volume $v$, expressed as constraint functionals:
\begin{mathletters}
  \label{eq:contconst}
  \begin{eqnarray}
    \lefteqn{
    \rho({\bf r}) =  \sum_{\alpha} \langle\delta({\bf r} - {\bf
      r}_\alpha)\rangle }\\\lefteqn{ = \sum_{j\ge 1} {j
      \choose 1}\left[
      \prod_{i=1}^{i=j} \int_v d {\bf r}_i \right] \delta({\bf r} - {\bf
      r}_1)p(j;{\bf r}_1,\ldots,{\bf r}_j)~,}\nonumber\\ 
    \lefteqn{
    \rho^2 g^{(2)}({\bf
      r}, {\bf r}')/2  =  \langle \sum_{\alpha>\gamma} \delta({\bf r} - {\bf
      r}_\alpha)\delta({\bf r}' - {\bf r}_\gamma)\rangle} \\ & =
    &\sum_{j\ge 2} {j \choose 2}\left[ \prod_{i=1}^{i=j} \int_v d {\bf
        r}_i \right] \delta({\bf r} - {\bf r}_1)\delta({\bf r} - {\bf
      r}_2)p(j;{\bf r}_1,\ldots,{\bf r}_j)~.\nonumber
  \end{eqnarray}
\end{mathletters}
where ${\bf r}$ and ${\bf r}'$ are positions inside the observation
volume $v$.  For a homogeneous fluid, the density will be uniform,
i.e., $\rho({\bf r})$ will be independent of position; and the
two-particle density distribution will be the radial distribution
function of the homogeneous system, $g^{(2)}({\bf r},{\bf r}')=g(|{\bf
  r}-{\bf r}'|)$.

In analogy to eq~\ref{eq:entropy}, we define an information entropy $\eta$
for continuous solvent positions and discrete occupancy
numbers,\cite{Stratonovich:JETP:55}
\begin{eqnarray} 
  \eta &\equiv& - \sum_j \left[ \prod_{i=1}^{i=j} \int_v
    d {\bf r}_i \right] p(j;{\bf r}_1,\ldots,{\bf r}_j) \ln[v^j
  j!p(j;{\bf r}_1,\ldots,{\bf r}_j)] .\nonumber\\
  \label{gentropy}
\end{eqnarray}
We include the factor of $v^j$ in the logarithm of eq~\ref{gentropy}
for dimensional consistency.  Maximizing this entropy functional under
the constraints of eq~\ref{eq:contconst} results in an expression for
the probability distribution $p(j;{\bf r}_1,\ldots,{\bf r}_j)$,
\begin{eqnarray}
  &&-\beta^{-1}\ln[v^j j!p(j;{\bf r}_1,\ldots,{\bf r}_j)] = \sum_{k=1}^j
  \omega^{(1)}({\bf r}_k)\nonumber\\&& + \frac{1}{2}
  \sum_{k,l=1}^j \beta\omega^{(2)}({\bf r}_k,{\bf r}_l) ~.  \label{pmin}
\end{eqnarray}
The Lagrange multipliers $\omega^{(1)}({\bf r}_k)$ and
$\omega^{(2)}({\bf r}_k,{\bf r}_l)$ are chosen such that the $p(j;{\bf
  r}_1,\ldots,{\bf r}_j)$ satisfy the constraint functionals
eq~\ref{eq:contconst}.  Incorporating additional $n$-particle
correlation information ($n>2$) is straightforward.  In a practical
implementation, one can subdivide the observation volume into a finite
number of volume elements.  The constraint functionals,
eqs~\ref{eq:contconst}, then
reduce to a finite number of constraints.

Note that the probability distributions $p(j;{\bf r}_1,\ldots,{\bf
  r}_j)$ have a Boltzmann-Gibbs structure, i.e., they are proportional
to an exponential of effective interactions $\omega^{(1)}({\bf r}_k)$
and $\omega^{(2)}({\bf r}_k,{\bf r}_l)$ in $k_{\rm B}T$ units, divided
by $j!$.  Summation and integration of the $p(j;{\bf r}_1,\ldots,{\bf
  r}_j)$ results in the familiar form of a grand-canonical partition
function for a microscopic volume $v$ embedded in a large bath of
solvent molecules.  The effective interactions entering the
Boltzmann-Gibbs factors in this grand-canonical partition function are
the Lagrange multipliers $\omega^{(1)}({\bf r}_k)$ and
$\omega^{(2)}({\bf r}_k,{\bf r}_l)$.  These effective interactions are
chosen to satisfy available information constraints, rather than being
derived directly from integrating out bath degrees of freedom.

$\omega^{(1)}({\bf r})$ is expected to be a
spatial constant approximately the negative chemical potential of the
solvent.\cite{Hummer:PRL:98} Similarly, $\omega^{(2)}({\bf r},{\bf
  r}')$ will be approximately an interatomic pair potential, except in
a small region near the surface of the subsystem.

\subsection{Some practical aspects of the information theory model}
Here, we provide details on the practical implementation the
IT model.  Before discussing specific details, we note
that the required moments of particle-number fluctuations can be
obtained {\em in situ} from simulation data.  This idea is most direct
for a uniform liquid.  The observation volume $v$ is
planted as a stencil in the simulation volume and the required moments
are extracted as averages over a set of simulation configurations.
For a uniform liquid the positioning of the observation volume is
irrelevant and many positions can be used simultaneously.  It is worth
noting further that since the occupation numbers can only be a finite number
of non-negative integers, only a finite number of binomial moments
$\langle { n \choose k} \rangle$ will be non-zero.  Similar procedures
apply to non-uniform systems such as the solvation structure near a
fixed solute, but the positioning of the observation volume is then
relevant.\label{sec:strat}

The observation volume may be subdivided
into several strata.  This introduces cross moments in the general
case and permits conceptually interesting possibilities for modeling
the probabilities.  As an example, consider the mean density in a
volume element external to a core region conditional on the
requirement that the core region be empty.  With such considerations,
the IT modeling naturally produces a prediction of the
hydration structure near a hydrophobic solute without special
geometrical limitations.  Such possibilities have scarcely been
studied so far, however.  For the calculation of the required moment
information, stratification is no particular problem but it may affect
the computational efficiency.  Note, that a stratified representation
can also be derived from coarse-graining the continuous IT
of section \ref{sec:cont}.

Next, we will illustrate several methods to calculate the
second moment of the particle-number fluctuations.  Then we will
describe numerical methods for the entropy maximization.

{\em Calculation of particle-number variances.} The second-moment
constraint eq~\ref{eq:momtwo} requires the calculation of a double
integral over the volume $v$ involving the distance-dependent pair
correlation function $g(r)$.  Such integrals can be evaluated using
Monte Carlo techniques.  The average of the pair correlation function,
$\langle g(|{\bf r}-{\bf r}'|)\rangle$ can be determined by placing
two points ${\bf r}$ and ${\bf r}'$ randomly inside the volume $v$
according to a uniform probability density.  This average is then
multiplied by the square of the volume $v$ to give an estimate of the
integral in eq~\ref{eq:momtwo}.

As suggested above, the variance and higher moments of particle-number
fluctuations can be determined directly from molecular simulations.
Unlike $p_0$, the mean and variance can be calculated accurately from
insertion also for large solutes.  Objects with volume $v$ (shape and
size) are inserted with random positions and orientations into water
configurations taken from an equilibrium molecular dynamics or Monte
Carlo simulation.  The mean and variance of the particle number $n$ is
then determined directly.

Volumes of certain shapes are amenable to analytical evaluation.  For
spherical volumes $v$ of radius $d$, Hill\cite{Hill:58} derived a
transformation of the six-dimensional integral to one-dimensional
form.  Such transformations can be accomplished by determining the
pair-distance distribution $P(R)$ well known from small-angle
scattering theory,\cite{Guinier:39}
\begin{eqnarray}
  \label{eq:pair}
  P(R) & = & \int_v d{\bf r} \int_v d{\bf r}' \delta(R-|{\bf r}-{\bf
  r}'|)~,
\end{eqnarray}
where $\delta(x)$ is Dirac's delta function.  This leads to the
following expression for the second moment:
\begin{eqnarray}
  \langle n^2\rangle & = & \langle n\rangle + \rho^2 \int dR\; P(R) g(R)~.
\end{eqnarray}
We can also use the three-dimensional analog to eq~\ref{eq:pair},
\begin{eqnarray}
  \label{eq:pairthree}
  P({\bf R}) & = & \int_v d{\bf r} \int_v d{\bf r}' \delta[{\bf R}
  -({\bf r}-{\bf r}')]~.
\end{eqnarray}
Fourier transformation relates $P({\bf R})$ to the form factor $S({\bf
  k})$ of the volume $v$,
\begin{eqnarray}
  \label{eq:formfac}
  \tilde{P}({\bf k}) & = & \int d{\bf r} P({\bf r}) e^{i {\bf
  k}\cdot{\bf r}} = |S({\bf k})|^2~,
\end{eqnarray}
where
\begin{eqnarray}
  \label{eq:S}
  S({\bf k}) = \int_v d{\bf r} e^{i {\bf k}\cdot{\bf r}}~.
\end{eqnarray}
The Fourier transform eq~\ref{eq:formfac} can be inverted as
\begin{eqnarray}
  \label{eq:Fourinv}
  P({\bf R}) & = & \frac{1}{(2\pi)^3} \int d{\bf k} \tilde{P}({\bf
  k}) e^{-i{\bf k}\cdot{\bf r}}~.
\end{eqnarray}
As a generalization of Hill's result for a single
sphere,\cite{Hill:58} we find for a collection of $N$ non-overlapping
spheres of radius $d$:
\begin{eqnarray}
  \label{eq:Nspheres}
  \lefteqn{P(R) = 4 \pi R^2 \left\{ \left[N \pi
  d^3\left(\frac{4}{3}-\frac{R}{d} +
  \frac{R^3}{12d^3}\right)\right]\theta(2d-R)\right.}\nonumber\\
&&\left.+\sum_{i=1}^{N-1}\sum_{j=i+1}^N \left[\pi d^3
  \left(\frac{4}{3}(1-x_{ij}) -
  \frac{{y_{ij}}^{3/2}-|R-r_{ij}|^3}{3r_{ij}Rd}\right.\right.\right.\\
\lefteqn{\left.\left.\left.
  +\frac{{y_{ij}}^{5/2}-|R-r_{ij}|^5}{60r_{ij}Rd^3}
\right)\right]\theta[R-(r_{ij}-2d)]\theta(r_{ij}+2d-R)\right\}~,\nonumber}
\end{eqnarray}
where $r_{ij}$ is the distance of the centers of spheres $i$ and $j$,
$x_{ij}=(R^2+{r_{ij}}^2-4d^2)/(2Rr_{ij})$, and
$y_{ij}=R^2+{r_{ij}^2}-2Rr_{ij}x_{ij}$.  Using this method, analytical
results can also be found for rotation ellipsoids with axes $a=b$ and
$c$.

{\em Maximizing the information entropy.} Maximization of the
information entropy eq~\ref{eq:entropy} is easily performed using
Lagrange multipliers for the constraints.  For constraints on mean and
variance, this leads to the expression eq~\ref{eq:pnmax} for $p_n$.
Lagrange multipliers $\lambda_0$, $\lambda_1$, and $\lambda_2$ are
then calculated such that the corresponding $p_n$ satisfy the moment
conditions eq~\ref{eq:moments}.  This corresponds to solving three
non-linear equations with three unknowns which can be accomplished, for
instance, using a Newton-Raphson method.  Alternatively, one can use
standard minimization packages, minimizing the squared differences of
the left and right-hand sides of the moment constraint equations.
This is consistent with the idea of minimizing the ``thermodynamic
potential''
\begin{eqnarray} f(\lambda_1,\ldots,\lambda_{k_{\rm max}}) & \equiv
& \ln \left[1+
\sum_{j=1}^{k_{\rm max}}{ \hat p}_j \exp \left(-\sum_{k=1}^{k_{\rm max}} \lambda_{k}
{j \choose k} \right) \right] \nonumber \\ & + &\sum_{k=1}^{k_{\rm max}}
\lambda_{k} \left\langle{j \choose k}\right\rangle~,  \label{pfunction} \end{eqnarray}
here expressed in terms of the binomial moment information and for a
general default model.

\subsection{Some pitfalls of the simplest models} 
\label{sec:pitfalls}
It is a helpful, heuristic view that the IT approach
studies a model grand partition function associated with a molecular
scale volume and involving effective interactions.  Because this may be an
unusual setting for the consequent statistical thermodynamic
calculations, conventional results are not guaranteed.  Here we note
some of the pitfalls that have been encountered for the simplest
models.  These discussions emphasize that the orientation for
development of this approach is to discover models that work.

{\em Non-separability of chemical potentials of distant sites within a
  Gaussian model.} Consider two spheres of exclusion radius $\lambda$
separated by a distance large compared to $\lambda$.  The excess
chemical potential of the two spheres {\em should be\/} two times that
of the individual spheres, $\mu^{\rm ex}(1,2)=2\mu^{\rm ex}$.  The mean
and variance of the particle-number fluctuations are also additive.
For the combined volume, we will have $m_2 = 2 m_1$ and
${\sigma_2}^2=2{\sigma_1}^2$, where $m_i$ and ${\sigma_i}^2$ are the
means and variances of the particle-number distributions of an
individual sphere $(i=1)$ and the two spheres $(i=2)$.  Following
the flat default model, we approximate $p_0$ by a Gaussian
form,\cite{Garde:PRL:96}
\begin{eqnarray}
  \label{eq:pzGauss}
  p_0 & \approx &
  \frac{e^{-{m_i}^2/2{\sigma_i}^2}}{\sqrt{(2\pi{\sigma_i}^2)}}~.
\end{eqnarray}
We then find that the corresponding chemical potential of an
infinitely-distant pair of spheres is not exactly that of two
individual spheres,
\begin{eqnarray}
  \label{eq:sumdiff}
  \mu^{\rm ex}(1,2) - 2\mu^{\rm ex} & \approx & - \frac{ k_{\rm B} T}{2} \ln (
  \pi {\sigma_1}^2 )~.
\end{eqnarray}
This non-additivity is a direct consequence of the Gaussian model and
can be simply repaired by insisting accurately on zero probability for
negative occupancies.  In fact, the definition of our problem here
requires that the occupancies be non-negative integers.

Additivity can be recovered by dividing the two sphere volume into two
strata: one for each sphere or, when the spheres overlap, the
individual strata can be defined by utilizing the plane of
intersection of the spherical surfaces as a bounding surface.
Suppose again that the spheres are far apart and that the probability
of occupancy of one sphere is independent of the occupancy of the
other.  The cross correlation between the numbers $n_1$ and $n_2$ of
solvent centers in the two non-overlapping strata is then given by
\begin{eqnarray}
  \label{eq:crosscorr}
  \langle n_1 n_2\rangle & = & \rho^2 \int_{v_1} d{\bf r}_1 \int_{v_2} d{\bf
    r}_2 g(|{\bf r}_1-{\bf r}_2|)\nonumber\\ & = &
  \rho^2 v_1 v_2 = \langle n_1\rangle
  \langle n_2\rangle~,
\end{eqnarray}
corresponding to {\em uncorrelated} particle-number fluctuations.  An
expanded IT model uses the joint probabilities
$p(n_1,n_2)$ of finding $n_1$ particles in $v_1$ and $n_2$ particles
in $v_2$.  Including the constraint eq~\ref{eq:crosscorr} derived from
stratification in that expanded IT model gives the
correct result that the hydration free energy of the two spheres is
the sum of the hydration free energies of the individual spheres.

The non-additivity in the simple model is caused by the possibility of
having a zero occupancy of the volume $v$ with a {\em negative} number
$-n$ of particles in sub-volume $v_1$ that is compensated by a
positive number $n$ of particles in sub-volume $v_2$.  By imposing
explicitly that each of the sub-volumes has zero occupancy when the
total occupancy is zero, the non-additivity is eliminated.  This
argument illustrates that the restriction of the probabilities $p_n$
to non-negative integers can be significant.  This example also
emphasizes that arguments well justified about the center of a
distribution can lead to fundamental errors when applied to the wings
of a distribution.

{\em Macroscopic limit.\/} For the flat default model, we can
calculate the limit of large solutes.  The variance in the particle
number is then given by the isothermal compressibility $\chi_T$,
\begin{eqnarray}
  \label{eq:comp} \frac{\langle n^2 \rangle - \langle
n\rangle^2}{\langle n\rangle} & = & \rho k_{\rm B} T \chi_T~
\quad\mbox{for $v\rightarrow\infty$}~.
\end{eqnarray}
For the flat default model and a spherical solute with exclusion radius
$\lambda$, we find that the chemical potential grows with the volume of
the solute,
\begin{eqnarray}
  \label{eq:vol} \frac{\partial\mu}{\partial \lambda} & \approx &
\frac{2\pi \lambda^2}{\chi_T}~,
\end{eqnarray}
where we used the Gaussian approximation eq~\ref{eq:pzGauss}.  The
flat default model predicts for the water density $\rho G(\infty)$ at contact
with a hard wall:
\begin{eqnarray}
  \label{eq:G} G(\infty) & = & \lim_{\lambda\rightarrow\infty} (4\pi
\lambda^2\rho k_{\rm B}T)^{-1} \frac{\partial\mu}{\partial \lambda}
\approx (2\rho k_{\rm B}T\chi_T)^{-1}~,
\end{eqnarray}
which for water under standard conditions results in
$G(\infty)\approx 8$.  The contact value for a flat wall, however, is
well known to be $G(\infty)=p/\rho k_{\rm B}T$, where $p$ is the
pressure.  If we use bulk density and pressure of water under standard
conditions, we obtain $p/\rho k_{\rm B}T\approx 7.4\times 10^{-4}$;
for saturated water vapor,\cite{Stillinger:73} we find $p/\rho k_{\rm
  B}T\approx 2\times 10^{-5}$.  Clearly, the contact value at a flat
wall, $G(\infty)$, would be grossly overestimated if the flat default
model were used in this macroscopic limit.

In the limit of large solutes we expect the multiphasic nature of
water to become important.  Stillinger\cite{Stillinger:73} pointed out
that the solvation of hard spheres much larger than a water molecule
can be quantified by considering that a molecularly thick vapor layer surrounds
the solute.  For large solutes, the formation of a
correspondingly large water-vapor bubble with subsequent insertion
into that bubble will cost less free energy than inserting directly
into the {\em liquid} phase.
The simplest IT model uses a flat or
Gibbs default model and does not account for this possibility, as
density fluctuations are of either Gaussian or Poisson-like character,
respectively.  More elaborate default models should however be able to
account for the possibility of vaporizing a certain small volume.  But
clearly, defining such models would require knowledge about the free
energy of microscopic vapor-bubble formation which is a task as
formidable as the calculation of solvation chemical potentials.

An indication of the large-solute effects can already be seen in
Figure~\ref{fig:sphere_pn}.  For solutes of size $d=0.3$ nm and larger,
we find that $p_1$ is depressed relative to the flat default model.
This means that it is relatively less likely to find an isolated water
molecule in a cavity than would be predicted from that simple model.
Based on the foregoing discussion, for solutes of increasing size we
expect that $p_0$ will be become larger than predicted from the flat
default model.

\subsection{Historical perspective}

{\em Scaled particle theory ideas.\/} The IT approach
can be viewed in the context of preceding studies of the insertion
probability $p_0$.  Mayer and Montroll\cite{Mayer:41} and later Reiss
et al.\cite{Reiss:59} expressed $p_0$ in terms of higher-order
correlation functions of the bulk fluid,
\begin{eqnarray}
  \label{eq:pzero_corr} p_0 & = & 1 + \sum_{m=1}^\infty \frac{(-1)^m
}{m!} \int_v d{\bf r}_1 \int_v d{\bf r}_2 \cdots \int_v d{\bf r}_m
\rho^m \nonumber\\&&\times g^{(m)}({\bf r}_1, {\bf r}_2, \ldots, {\bf r}_m)~,
\end{eqnarray}
where $\rho^m g^{(m)}$ is the $m$-body joint density for solvent
centers (here, water-oxygen atoms).  These are standard combinatorial
results, more frequently seen in forms such as\cite{Riordan:78}
\begin{eqnarray}
  \label{eq:facmom} p_0 & = & 1 + \sum_{m=1}^\infty (-1)^m\left\langle
{n\choose m}\right\rangle~,
\end{eqnarray}
and\begin{eqnarray}
p_j =
\sum_{m=0} (-1)^m {j+m \choose j}\left\langle {n \choose j+m} \right\rangle~,
\label{eq:riordan} \end{eqnarray}
where $n$ is the number of solvent centers within $v$.
The first of these can be derived directly by noting that $p_0 =
\langle (-1+1)^n\rangle$ where $(-1+1)^n$ is an indicator function
that is one for $n=0$ and zero otherwise.  These sums truncate sharply
if only a finite maximum number of particles can be present
in the observation volume.  They are of practical value for small
volumes where the maximum number of solvent centers that can possibly
occupy that volume is small.  For larger volumes, large terms of
alternating sign appear in the expansion eq~\ref{eq:facmom} that make
it difficult to use estimated values for the factorial moments in this
linear fashion.

Equating the excess chemical potential $\mu^{\rm ex}$ to the
quasi-static work of creating a cavity results in an expression for
$p_0$ in terms of the contact value $G(\lambda)$ of the solvent
density at the surface of the exclusion volume $v$.  For a spherical
exclusion volume of radius $\lambda$ we have:\cite{Reiss:59,Reiss:65}
\begin{eqnarray}
  \label{eq:SPT-G}
  \mu^{\rm ex} & = & -k_{\rm B}T \ln p_0 =  4\pi \rho k_{\rm B}T
  \int_0^d G(\lambda) \lambda^2 d\lambda~.
\end{eqnarray}
This result is formally exact for
a hard-sphere solute and establishes that $ 4\pi \rho k_{\rm B}T
G(\lambda) \lambda^2$ is the compressive force exerted by the solvent
on a hard spherical solute of diameter $\lambda$.  If the solvent is
considered to contain a hard core itself, there is a maximum number of
solvent particles $n_{\rm max}$ that can occupy a certain volume $v$.
This results in useful, exact expressions for the case of $n_{\rm
max}=1$ and 2.  For larger solutes, the expected macroscopic solvation
behavior\cite{Pierotti:63,Stillinger:73}
\begin{eqnarray}
  \label{eq:SPTass}
  \mu^{\rm ex} \sim  p (4\pi \lambda^3/3) +  \gamma (4 \pi \lambda^2)~,
\end{eqnarray}
is used as a basis for extrapolation.  Here $p$ is the pressure and
$\gamma$ is the liquid-vapor surface tension.  In the intermediate
molecular region, the approximate asymptotic form is connected
smoothly to the exact curve for small
solutes.\cite{Pierotti:63,Stillinger:73}

{\em Computer simulation studies.} The cavity formation probability
$p_0$ lends itself to direct computer simulation studies by
test-particle insertion.  Spherical volumes $v$ in water were studied
extensively using computer simulations.
\cite{Pratt:PNAS:92,Pohorille:JACS:90,Palma,Guillot:JCP:93,%
  Forsman:94,Beutler:95:a} Similar insertion methods were also used to
study polymer solubilities.\cite{Stamatopoulou:JCP:98} The simulation
studies of Pohorille and
Pratt\cite{Pratt:PNAS:92,Pohorille:JACS:90,Palma} clarified a number
of interesting speculations on hydrophobicity and provided the first
discriminating tests of theoretical models for G($\lambda$).

\if T\TS \FIGfour \fi

One hypothesis that has sparked interest in recent years is the idea
that the low solubility of inert gases in liquid water is due to the
small size of the water molecule and, consequently, the
``interstitial'' cavities
will be smaller in water than in coexisting organic liquids of common
interest.\cite{Lee:85,Lee:91} Since the present work deals directly with
such cavities, it is straightforward to test that idea.
Figure~\ref{fig:bklee}\nocite{Palma} shows distributions\cite{Palma}
of the distances from an arbitrary point to the nearest interaction site
for three cases of interest: liquid n-hexane, liquid water, and an
ideal gas with water density.
Reasonable and customary values are used for the van der Waals
radii of the solvent interaction sites in each case.  Figure~\ref{fig:bklee}
shows that the most probable cavity radii are quite small in all those
cases as is expected for liquids.  Also, the difference in the
most probable cavity size between liquid n-hexane and liquid water is not
large, and smaller than the corresponding difference between liquid
water and an ideal gas with water density.  The fact that
the differences seen here are slight is, presumably, defined by
the fact that the most basic units considered in n-hexane are the
methyl and methylene groups. These are not so different in size from a
water molecule.  In contrast, simple equation of state models might
treat the n-hexane molecule as a sphere of substantially larger size.
It can also be noted that on a packing fraction basis, typical organic
liquids are denser than liquid water.\cite{Pohorille:JACS:90}
        
One feature in Figure~\ref{fig:bklee} that does distinguish water is
the breadth of the distributions.  The distribution of the nearest
solvent site neighboring an arbitrary position in the liquid is {\em
  narrower\/} in liquid water.  The widths of these distributions are
independent of the van der Waals radii
assigned to the solvent sites.  The interpretation is
that, compared to other liquids, water is ``stiffer'' when larger
cavities must be opened.
\label{sec:size}

\if T\TS \FIGfive \fi

This comparative stiffness can be analyzed further by determining the
contact functions $G(\lambda)$ for water and hydrocarbon liquids and
comparing the compressive force that such solvents exert on a hard
spherical solute.  As shown in Figure~\ref{fig:Gforce}\nocite{Pratt:PNAS:92},
water exerts a higher compressive force on such a
solute than typical hydrocarbon solvents.  It is thus an accurate view
on molecular-scale hydrophobic effects that water ``squeezes-out''
hydrophobic solutes.\cite{Richards:SA:91,Haymet:Fara:96}

{\em Field theory models.} A perspective related to the IT
 approach pursued here has been adopted before by
Chandler.\cite{Chandler:PRE:93} Equilibrium fluctuations of the water
density are approximated by a Gaussian model.  The partition functions
are then calculated for the two systems with and without a cavity of
volume $v$ present, using a field-theoretic method.
Chandler then shows that this continuous
Gaussian fluctuation model leads to the Pratt-Chandler integral
equation theory.\cite{Pratt:JCP:77} For sufficiently large volumes, a
Gaussian field model is expected to reproduce accurately the $p_n$ if
the continuous probability densities of particle numbers are mapped to
discrete values of $n$ and the negative values of $n$ are disregarded.

\if T\TS \FIGsix \fi

\section{Results for elementary hydrophobic effects}
\subsection{Hydration of simple spherical solutes}
The probabilities $p_n$ of observing exactly $n$ oxygen centers inside
spherical volumes, shown in Figure~\ref{fig:sphere_pn}, are reproduced
accurately using IT with a flat default model.  The
corresponding excess chemical potentials of hydration of those
solutes, $\mu^{\rm ex}=-k_{\rm B}T\ln p_0$, are shown in
Figure~\ref{fig:sphere_mu}.\cite{Hummer:PNAS:96} As expected,
$\mu^{\rm ex}$ increases with increasing cavity radius.  The agreement
between IT predictions and computer simulation results
is excellent over the whole range $d\leq 0.36$ nm accessible to direct
simulation calculations of $p_0$.

\if T\TS \FIGseven \fi

\subsection{Methane-methane potential of mean
  force (PMF)} We now focus on the simplest model of hydrophobic
interactions exemplified by the free energy of bringing together two
methane-size cavities in water.
\cite{Pratt:JCP:77,Pangali:79,Watanabe:86,Smith:JCP:93,Wodak:93,%
  Head-Gordon:JACS:95,Garde:96:b,Ludemann:97} This free energy profile
defines a PMF.  We obtain the cavity
contribution to the methane-methane PMF by calculating the excess
chemical potential, $\mu^{\rm ex}(r)$, of a cavity volume described by
two spheres each of radius $d$ and separated by a distance $r$.  The PMF
is then defined as $W(r)=\mu^{\rm ex}(r) - \lim_{s\rightarrow\infty}
\mu^{\rm ex}(s)$.  This definition guarantees that $W(r)\rightarrow 0$
for $r\rightarrow\infty$, avoiding the problem of non-additivity
discussed above for non-stratified volumes.

Figure~\ref{fig:PMF} shows a comparison of the cavity PMF calculated
using IT with that obtained from explicit simulations
by Smith and Haymet.\cite{Smith:JCP:93} The flat default model has
been used here and throughout this paper, unless otherwise noted.
Also, an exclusion radius of $d=0.33$ nm has been used for
methane-size cavities corresponding to a distance where the
methane-water oxygen radial distribution function reaches a value of
1.0 in commonly used models.\cite{Hummer:96:a,Jorgensen:84} The direct
interaction between the two methane molecules has been subtracted from
the simulation PMF to get approximately the cavity contribution to the
methane-methane PMF.

\if T\TS \FIGeight \fi

In agreement with the simulation data, the PMF calculated using
IT shows features characteristic of hydrophobic
interactions.  A free energy minimum is observed corresponding to
overlapping hard spheres.  That deep minimum is separated by a barrier
from a second minimum at a distance of about 0.7~nm. The second
minimum corresponds to solvent-separated pairs of methane-size
cavities.  In addition, we observe a third shallow minimum which was
also seen in computer simulations by L\"{u}demann et
al.\cite{Ludemann:97} The equilibrium between the solvent-separated
and contact minimum as a function of pressure provides insights into
pressure denaturation of proteins, as described in section
\ref{sec:pressure}.

\subsection{Solvent contributions to the torsional equilibrium of
  butane} As a final example of a simple hydrophobic effect,
Figure~\ref{fig:butane} shows the cavity contribution to the torsional
PMF for $n$-butane in water.  IT calculations are
compared with explicit computer simulations by Beglov and
Roux.\cite{Beglov:94} We find that the the cavity PMF favors the
compact cis ($\phi=0$) structure over the extended trans structure by
about 1.8 kJ/mol.  The gauche structure ($\phi=\pi/3)$ is favored over
trans by about 0.7 kJ/mol.  Those differences and the overall cavity
PMF for the torsional isomerization of butane are in excellent
agreement with the simulation data.

\if T\TS \FIGnine \fi

\section{Temperature dependence of hydrophobic hydration}
\subsection{Solvation chemical potentials}
The hydrophobic effect is often associated with characteristic
temperature dependences.\cite{Baldwin:PNAS:86,Dill:S:90} One of the more
astonishing observations is that the entropies of transfer of
nonpolar molecules from gas phase or a nonpolar solvent into water
converge at a temperature of about 400~K to approximately zero entropy
change.  Similar behavior was also seen in microcalorimetry
experiments on unfolding of several globular
proteins.\cite{Privalov:APC:79,Privalov:APC:88,Makhatadze:APC:95}
This analogy to the transfer data supported the hydrophobic-core model
of protein folding: during unfolding predominantly nonpolar residues
are transferred from a mostly nonpolar protein core into an aqueous
environment.

We have used the IT model to clarify the molecular
origin and the quantitative details of this striking
entropy-convergence behavior.\cite{Garde:PRL:96} Computer simulations
were carried out to calculate the water-oxygen radial distribution
function $g(r)$ at several temperatures along the experimental
saturation curve of water.  The $g(r)$'s together with the
corresponding water densities are used as input in the IT
model to calculate chemical potentials of hydration for hard
solutes.  Solute sizes of $d=0.28$, 0.31, 0.33, and 0.345 nm were
considered, corresponding approximately to neon, argon, methane, and xenon, 
respectively.

Figure~\ref{fig:muT} shows the chemical potentials for each of the
spherical solutes as a function of temperature $T$, for $T$ between
275 and 625~K.  Also shown are the results of direct simulation
calculations using test-particle insertion together with
representative statistical errors.  We find excellent agreement
between simulation and theory for the chemical potentials over the
whole temperature range.  The chemical potentials are approximately
parabolic except at the highest temperatures, and the curves for
different solute sizes are shifted vertically with the maximum at
about 400 K.

\if T\TS \FIGten \fi

We obtain a solvation entropy by taking the derivative of the chemical
potential along the saturation curve,
\begin{eqnarray}
  \label{eq:ent}
  S_{\rm sat} & = & - \left( \frac{\partial\mu^{\rm ex}}{\partial T}
  \right)_{\rm sat}~.
\end{eqnarray}
Additional equation-of-state contributions to the standard solvation
entropy are negligible, estimated to be less than 1 and 10
J/(mol K) for temperatures $T<450$ and $T<550$~K, respectively.
Figure~\ref{fig:ent} shows the temperature dependence of the entropy
$S_{\rm sat}$ for the different solutes.  The entropies are large
and negative at room temperature for all the solutes and decrease
in magnitude with increasing temperature.  The temperature dependence
of entropies is approximately linear with slopes increasing with the
increasing solute size.  The resulting heat capacity,
\begin{eqnarray}
  \label{eq:heat}
  C_{\rm sat} & = & T \left( \frac{\partial S_{\rm sat}}{\partial T}
  \right)_{\rm sat}~,
\end{eqnarray}
is large and positive (approximately 40 cal mol$^{-1}$ K$^{-1}$ for a
methane-size cavity).  Moreover, the entropies converge at about 400~K
to approximately zero entropy, although at closer inspection the
temperature range of the convergence region is several 10 K and the
entropy is not exactly zero at convergence.  To understand the origin
of this approximate convergence we determine in the following the
factors entering into the IT model.

\if T\TS \FIGeleven \fi

\if T\TS \FIGtwelve \fi

\subsection{Origin of entropy convergence in hydrophobic hydration and
  protein folding}

To quantify the different factors contributing to the entropy
convergence,\cite{Garde:PRL:96} we use the approximate Gaussian
representation eq~\ref{eq:pzGauss} for $p_0$ with mean $m=\langle
n\rangle=\rho v$ and variance $\sigma^2 = \langle n^2\rangle - \langle
n\rangle^2$.  This results in an estimate of the chemical potential,
\begin{eqnarray}
  \label{eq:muex}
  \mu^{\rm ex} &\approx& T\rho^2\{k_{\rm B} v^2/2\sigma^2\} + T \{k_{\rm B}
  \ln(2\pi \sigma^2)/2\}~.
\end{eqnarray}
The second term is smaller than the first, and depends only
logarithmically on the solute size.  The solvation chemical potential
may therefore be lowered by lowering the product of temperature and
density $T\rho^2(T)$, or by increasing the particle-number fluctuations
$\sigma^2$.  The temperature dependence of $T\rho^2(T)$ along the
saturation curve is non-monotonic.  On the other hand, we find that
the variance $\sigma^2(T,v)$ changes only little over the range of
temperatures in Figure~\ref{fig:muT}.\cite{Garde:PRL:96}
Accordingly, we can approximate
the chemical potential as
\begin{eqnarray}
  \label{eq:muxy}
  \mu^{\rm ex} & \approx & T\rho^2(T)x(v) + T y(v)~,
\end{eqnarray}
where the functions $x(v)=k_{\rm B} v^2/2\sigma^2$ and $y(v)=k_{\rm B}
\ln(2\pi \sigma^2)/2$ depend only on the solute volume, not on the
temperature.  The contributions of the various terms to the entropy
convergence are illustrated schematically in
Figure~\ref{fig:ent_schem}.  The term $Ty(v)$ in eq~\ref{eq:muex}
is generally small.  If we neglect $Ty(v)$, the solvation entropies
eq~\ref{eq:ent} converge exactly at zero entropy for solutes of
different size, and the temperature of convergence corresponds to 
the maximum of $T\rho^2(T)$ along the saturation curve.  When the
second term $Ty(v)$ is included, the entropy at convergence is 
different from zero, but the convergence is still exact.  However,
both $x(v)$ and $y(v)$ show weak temperature dependences.  When 
those are considered as well, the resulting entropy convergence is 
no longer exact but occurs over a temperature range that is several 
10 K wide for the solutes considered here.

Our analysis shows that the main factors leading to an entropy
convergence are derived from the properties of pure water.  They are:
(1) the non-monotonic behavior of $T\rho^2(T)$ along the saturation
curve and (2) the weak temperature dependence of the particle-number
fluctuations $\sigma^2(T,v)$ for solute excluded volumes.  The
macroscopic analog of $\sigma^2$ is the isothermal compressibility
$\chi_T$.  For macroscopic volumes $v$, eq~\ref{eq:comp} relates the
two, $\sigma^2 = \rho^2 v k_{\rm B}T\chi_T$.
Figure~\ref{fig:comp}\nocite{Rowlinson:Swinton:82} compares the
isothermal compressibility of water and several organic solvents up to
their critical point as a function of temperature.  Indeed, we find
that water shows far weaker temperature variations of $\chi_T$ than
the organic solvents, which can be seen as the origin of the entropy
convergence.  In the macroscopic limit, the variance $\sigma^2 =
\rho^2 v k_{\rm B}T\chi_T$ is a product of $T\rho^2(T)$ with a maximum
at about 440 K and $\chi_T$ with a minimum at about 320 K.  This
results in only small variations of $\sigma^2$ in the range of
temperatures between 273 and 420 K.

\if T\TS \FIGthirteen \fi

\section{Pressure denaturation of proteins}
\label{sec:pressure}
\subsection{Denaturation by water transfer into the protein interior}
Proteins can be denatured by pressures of typically a few kilobar.
\cite{Brandts:70,Zipp:73,Heremans:82,Weber:83,Silva:93,%
  Jonas:94,Royer:95,Silva:97} This pressure-induced unfolding has been
a long-standing puzzle in our understanding of protein stability.  The
``liquid-hydrocarbon model'' of protein
unfolding\cite{Baldwin:PNAS:86} explains unfolding as a transfer of
predominantly hydrophobic residues from the protein interior into the
aqueous solvent.  Indeed, this model accurately describes the
thermodynamics of temperature denaturation, in particular, the
convergence of the entropies of unfolding.  However, as
Kauzmann\cite{Kauzmann:87} pointed out, such a description fails to
explain the pressure denaturation of proteins.  In particular, the
volume changes associated with transferring hydrocarbons from a
nonpolar phase into water exhibit behavior exactly opposite to that
upon pressure unfolding as a function of pressure.  The volume change
of hydrocarbon transfer is negative at low pressure and positive at
high pressure;\cite{Sawamura:JPC:89}
the volume change of protein unfolding is positive at
low pressure and negative at high pressure.\cite{Kauzmann:87}

To explain the pressure denaturation of proteins, we invoke additional
information about the structural properties of pressure-denatured
proteins.  Nuclear magnetic resonance (NMR) measurements reveal
considerably more structural organization in pressure-denatured
proteins compared to that in temperature-denatured
proteins.\cite{Zhang:95} These and other
experiments\cite{Heremans:82,Weber:83,Silva:93,Jonas:94} indicate that
the pressure-denatured proteins are relatively more compact than
temperature-denatured proteins.  Accordingly, we depart from the
commonly used model explaining protein unfolding thermodynamics as a
transfer of nonpolar residues from the protein interior into solution.
Instead, we describe pressure denaturation as a transfer of water
molecules from the aqueous phase into the hydrophobic core of a
protein.

To quantify this inverted ``liquid-hydrocarbon model,'' we focus on
the stability of hydrophobic aggregates formed in water as a function
of pressure.  Aggregate formation of spherical, methane-size solutes
serves as an idealized model of the stability of protein hydrophobic
cores.  The free energy of forming such aggregates can be expressed in
terms of two-body and higher order PMF's between the methane-size
solutes in water.  In the following, we will study the pair and
three-body interactions of methane-like solutes in water which control
the stability and formation of larger aggregates.

\subsection{Effect of pressure on hydrophobic interactions}
Figure~\ref{fig:Wp}\nocite{Hummer:PNAS:98} shows the PMF's $W(r)$
between two methane-like particles in water for pressures up to about
7 kilobar (700 MPa) calculated using IT.\cite{Hummer:PNAS:98}
The PMF's exhibit two minima, a contact
minimum at about 0.4 nm distance and a solvent-separated minimum at a
distance of about 0.7 nm.  The two minima are separated by the
desolvation barrier.  The PMF's are normalized at the
solvent-separated minimum to illustrate that the contact minimum
loses importance with increasing pressure.  The free energy of the
contact minimum increases relative to the solvent-separated minimum by
about 0.9 kJ/mol when the pressure is increased from 1 bar to 7
kilobar.  In addition, the desolvation barrier between the two minima
also increases with increasing pressure.  Similar pressure
destabilization of contact configurations was observed for three
methane molecules in contact.\cite{Hummer:PNAS:98}

The effect of pressure on pair PMF's between two methane molecules has
also been studied by computer simulation.\cite{Payne:97} These simulation
calculations showed a pressure destabilization of contact pairs that
is comparable in magnitude to the IT calculations.  In
addition, simulation calculations also showed the formation of methane
aggregates in water at low pressures that dissolved at high
pressures.\cite{Wallqvist:JCP:92}

The pressure destabilization of hydrophobic contact configurations can
be understood from the following simplified model: At low pressures,
the interstitial space between two large nonpolar solutes is
energetically unfavorable for water
molecules.\cite{Hummer:PRL:98,Wallqvist:JPC:95:b} As the pressure
increases, the space between the nonpolar solutes is more likely to be
occupied by water, increasing the importance of solvent-separated
configurations relative to contact configurations.  A more detailed
analysis indeed shows that the destabilizing effect of pressure on
hydrophobic contacts increases with increasing size of the two
nonpolar solutes.\cite{Hummer:PNAS:98}
  
\subsection{Stability of proteins}
We have shown that the stability of hydrophobic aggregates decreases
with increasing pressure as a result of water penetration.  When
applied to the hydrophobic-core model of proteins, we predict that
pressure denatures proteins by swelling.\cite{Hummer:PNAS:98} A
swelling mechanism has also been proposed for urea-induced protein
denaturation.\cite{Wallqvist:JACS:98} Under sufficiently high
pressures water molecules will intercalate into the hydrophobic
protein interior.  The resulting structures of pressure-denatured
proteins are predicted to be more ordered and more compact than those
of temperature-denatured proteins.  Recent small-angle scattering
experiments\cite{Panick:98} indeed showed that the pressure-denatured
staphylococcal
nuclease protein has a considerably smaller radius of gyration of
$R_g\approx 3.5$ nm compared to the temperature-denatured protein with
$R_g\approx 4.6$ nm.

\subsection{Formation of clathrate hydrates}

This physical picture of pressure denaturation of proteins is relevant
also for the formation of clathrate hydrates, crystalline compounds
forming at elevated pressures from non-polar molecules and water.  As
solid deposits in gas pipelines, clathrate hydrates cause problems in
natural gas transmission.  Methane-water clathrate hydrates, for
instance, form at pressures above about 44 MPa at a temperature of 298
K, with a stoichiometry of about 6-7 water molecules per methane and a
methane-methane nearest-neighbor distance of about 0.62-0.74
nm.\cite{Sloan:90} At lower pressures, the two fluid phases separate.
Formation of the clathrate structure is consistent with stabilizing
the solvent-separated minimum in the methane-methane PMF relative to
the contact minimum, as predicted from the analysis of pressure
effects on hydrophobic interactions.  In addition, the present theory
predicts that larger hydrocarbons require lower pressure for clathrate
formation,\cite{Hummer:PNAS:98} in agreement with the experimental
observations that led to the gas-gravity method of phase
determination.\cite{Sloan:90}

\if T\TS \FIGfourteen \fi

\subsection{Pressure effects on protein folding kinetics}
We noted before that increasing pressure results in a higher
desolvation barrier between the contact configuration and the
solvent-separated minimum.  Figure~\ref{fig:Wbarr} shows the barrier
heights $\Delta W^{\ddag}_{f}$ and $\Delta W^{\ddag}_{u}$ from the
solvent-separated and contact minimum, respectively, as a function of
pressure.  Thus, $\Delta W^{\ddag}_{f}$ and $\Delta W^{\ddag}_{u}$ are the 
barriers for forming and breaking hydrophobic contact configurations, corresponding 
to ``folding'' and ``unfolding'' reactions, respectively.  We find 
that $\Delta W^{\ddag}_{f}$ and
$\Delta W^{\ddag}_{u}$ both increase approximately linearly with 
increasing pressure.  Activation volumes defined as 
$\Delta v^{\ddag}_{f/u} = \partial \Delta W^{\ddag}_{f/u}/\partial p$
are both positive, $\Delta v^{\ddag}_{f}=3.8$ ml/mol and $\Delta
v^{\ddag}_{u}=1.6$ ml/mol,  for folding and unfolding reactions,
respectively.  Increasing pressure is thus expected to
slow down both the ``folding'' and the ``unfolding'' reactions.

Indeed, experimental studies of the pressure-dependent folding and
unfolding kinetics of the protein staphylococcal nuclease showed a
reduction of both rates with increasing pressure.\cite{Vidugiris:95} 
The experimentally observed activation volumes for folding and unfolding
of staphylococcal nuclease are $\Delta V^{\ddag}_{f}=92$ ml/mol and
$\Delta V^{\ddag}_{u}=20$ ml/mol, respectively.  From the ratio of 
overall experimental activation volumes and theoretical activation 
volumes calculated per hydrophobic contact, we can estimate the 
number of hydrophobic contacts broken in the folding transition 
state.  For staphylococcal nuclease we estimate that number to be between 
10 and 25, in good agreement with the predictions of energy landscape 
theory.\cite{Hummer:PNAS:98,Bryngelson:95}

\if T\TS \FIGfifteen \fi

\section{Salt effects on hydrophobic hydration}
Addition of salts to water generally reduces the solubility of
nonpolar solutes.  The Hofmeister series\cite{Baldwin:BJ:96} ranks
salts, in part, according to this reduction of the solubility of 
nonpolar compounds.  Here, we do not attempt to provide a complete
description of salt effects on hydrophobic hydration.  Instead, we
present some preliminary results for the solubility of small
hard-sphere solutes in a solution of sodium-chloride (NaCl) in water.
In particular, we compare results for NaCl concentrations of 1, 3, 
and 5 mol/l as well as pure water.  The three salt solutions were 
studied using molecular dynamics simulations extending over 1 
nanosecond each.  For simulation details, see 
Ref.~\onlinecite{Hummer:JPCM:94}.

Figure~\ref{fig:salt} shows the excess chemical potential of spherical
solutes as a function of their exclusion radius $d$ with water.  The
solute-solvent interactions are modeled as those of hard spheres with
radii $R_{\rm w}=0.132$ nm of water, $R_{\rm Na}=0.085$~nm of Na$^+$,
and $R_{\rm Cl}=0.18$~nm of Cl$^-$.  The exclusion radii with water,
Na$^+$, and Cl$^-$ are $d=R+R_{\rm w}$, $R+R_{\rm Na}$, and $R+R_{\rm
  Cl}$, respectively, where $R$ is the solute hard-sphere radius.  We 
find that adding salt indeed increases
$\mu^{\rm ex}$ and thus decreases the solubility for a solute of a
given size.  For a methane-size solute ($d=0.33$ nm), the excess
chemical potential increases by about 10 kJ/mol from about 34 to 44
kJ/mol when increasing the NaCl concentration from 0 to 5 mol/l.

To describe the salt dependence using the IT approach,
we extend our formulation to solvent mixtures.  For ionic solutes, the
strong interactions between water and ions give rise to the formation
of compact hydrated ions which result in non-trivial density
fluctuations near ions.  Therefore, instead of trying to model the
full distribution of particle number fluctuations $p(n_{\rm W},n_{\rm
  Na},n_{\rm Cl})$, we focus on a more tractable subset, which
contains the probability of the event we are ultimately interested 
in, $p(n_{\rm W}=0,n_{\rm Na}=0,n_{\rm Cl}=0)$.  Further, we factorize 
this probability into an ionic part and a conditional water part,
\begin{eqnarray}
  \label{eq:pzerofact}
  &&p(n_{\rm W}=0,n_{\rm Na}=0,n_{\rm Cl}=0) = p(n_{\rm
    Na}=0,n_{\rm Cl}=0)\nonumber\\
  &&  \quad\times p(n_{\rm W}=0|n_{\rm Na}=0,n_{\rm Cl}=0)~,
\end{eqnarray}
where $p(n_{\rm W}=0|n_{\rm Na}=0,n_{\rm Cl}=0)$ is the conditional
probability of having no water-oxygen atoms in the water exclusion
volume, given that there are no ions in the respective ion exclusion
volumes.  Such a factorization combined with accurate simulation data
provides insights into salt effects on hydrophobic solubilities at a
molecular level.  The contributions to the solute chemical potential,
$-k_{{\rm B}}T \ln p(n_{\rm Na}=0,n_{\rm Cl}=0)$ and $-k_{{\rm B}}T
\ln p(n_{\rm W}=0|n_{\rm Na}=0,n_{\rm Cl}=0)$ both increase with
increasing NaCl concentration.  The increase in the salt contribution
is primarily due to an increase in the mean of the $p(n_{\rm
  Na},n_{\rm Cl})$ distribution.  Interestingly, the mean of $p(n_{\rm
  W}|n_{\rm Na}=0,n_{\rm Cl}=0)$ is approximately independent of salt
concentration.  It is the decrease in the variance of this conditional
water distribution with increasing NaCl concentration that leads to a
decrease in $p(n_{\rm W}=0|n_{\rm Na}=0,n_{\rm Cl}=0)$.  The direct
salt effect of the reduction in free volume due to overlap with salt
ions and the indirect salt effect of changing the water structure
account for roughly one half each of the total increase in solute
chemical potentials.

\section{New directions: Cavity expulsion and area laws}
We mentioned in section~\ref{sec:pitfalls} that for macroscopic
solutes physical effects intrude that are not captured in the simplest
IT models.  This is a consequence of the multiphasic
nature of water.  In the limit of large solutes, the solvation
chemical potential is largely determined by the free energy of forming
a vapor bubble of appropriate size.  The free energy of forming large
bubbles can be approximated by the product of surface area and
liquid-vapor surface tension, since the contribution from the
pressure-volume work is small for water.\cite{Stillinger:73}
Accordingly, we expect that the solvation chemical potential grows
with the surface area for large solutes.  Stillinger\cite{Stillinger:73} gave a
justification of this based on the water structure near a large hard
solute.  The contact-value theorem gives, in the limit of a flat hard
wall, a water density at contact of about $7.4 \times 10^{-4}$
times the bulk density of water.  Stillinger argued that this
``dewetting'' implies the presence of a molecularly thick vapor layer
near a hard wall, and as a result, area laws with surface tension
coefficients similar to that of the liquid-vapor surface tension of
water.

In an attempt to bridge the gap between microscopic and macroscopic 
solvation behavior, we have recently developed a quantitative 
description of the solvation of nonpolar solutes as a function of
their size.\cite{Hummer:PRL:98} The water structure near 
spherical\cite{Hummer:PRL:98} and non-spherical Lennard-Jones
solutes\cite{Garde:PRE:96,Ashbaugh:96} was found to be largely insensitive 
to the details of the solute-water interactions except for the solute 
size, motivating the use of perturbation theory.  For spherical 
solutes, we find that the radial distribution function of water 
around the solute can be described accurately by
\begin{eqnarray}
  \label{eq:g_perturb}
  g_{sw}(r) & = & \exp[-\beta u_{\rm rep}(r) - \beta \omega(r) + C(r)]~,
\end{eqnarray}
where $u_{\rm rep}(r)$ is the repulsive solute-water interaction from,
e.g., Weeks-Chandler-Anderson separation;\cite{Weeks:71} $C(r)$ is a
renormalized attractive interaction; and $\omega(r)$ represents the
free energy of a test water molecule that does not interact with the
solute as a function of the distance $r$ from the
solute,\cite{Hummer:PRL:98} thus containing the non-trivial solvent
contributions.  Interestingly, $\omega(r)$ is directly related to the
one-body Lagrange multiplier $\omega^{(1)}({\bf r})$ of the continuous
IT eq~\ref{pmin}.  We find that $\omega(r)$ is
dominated by a contribution $\omega_0(r-r_0)$ that is identical for
various solutes except for a simple radial translation with the solute
size $r_0$.  The remainder $\Delta\omega (r-r_0)$, however, changes
non-trivially with the solute size, acting as a cavity-expulsion
potential for the test water molecule.

The cavity-expulsion potential, $\Delta\omega(r-r_0)$, quantifies
primarily the loss of energetic interactions of a test water molecule
as it crosses the solute-water interface from the water phase towards
the center of the solute.  For small solutes, $\Delta\omega(r-r_0)$ is
negligible.  As the solute size increases, this loss of interactions
becomes important, reaching $-\mu^{\rm ex}_w$, the negative excess
chemical potential of water, in the limit of large solutes. For
solutes of diameters more than about 0.5~nm the cavity expulsion
results in a slight depression in the water density at the
solute-water interface.  We find that this ``weak dewetting'' with
increasing solute size is sufficient to give an approximate
surface-area dependence of the solvation chemical potential, which
would otherwise be dominated by solute volume
terms.\cite{Hummer:PRL:98}

\section{Concluding remarks}

How is water different from hydrocarbon liquids as a solvent for
nonpolar solutes?  This is the foremost question asked of theories of
hydrophobic effects and several of the points above constitute relevant parts
of the answer.  It is appropriate therefore to ask this question more
specifically and to organize features of the answer that are provided
by the IT models considered here. One basic aspect of
the answer is straightforward and on that score we do not contribute
anything further: water is different because of the
mechanical potential energies of the interactions among water
molecules, including hydrogen bonding, orientational specificity, and
cooperativity.
        
In the context of solvation thermodynamics, the question how water
differs from other solvents solicits information about a particular
structural pattern of the solvent that might be principally
responsible for hydrophobic hydration free energies.  For example,
clathrate-like structures have been invoked to explain hydrophobic
hydration phenomena.\cite{Head-Gordon:PNAS:95} The answer given by the
IT model is that water differs from common hydrocarbon
solvents, in the first place, because the distribution of pairs of
exclusion spheres, oxygen-oxygen sites in water, is
distinctive and has a distinctive temperature dependence.  In the
second place, the isothermal compressibility of water, that can be
obtained by integration of the oxygen-oxygen distribution, is
insensitive to temperature compared to the same property of organic
solvents.  Since the minimum in the temperature dependence of the
isothermal compressibility is very distinctive, this answer is
remarkable and remarkably simple.

From our analysis of pressure effects on hydrophobic interactions and
protein stability, we can deduce further peculiarities of water.
Increasing the pressure in water shifts the balance from attractive
and directional hydrogen-bonding forces to repulsive and isotropic
packing forces between water molecules.\cite{Stillinger:74} This
results in ``energetic frustration'' of water molecules in the water
phase,\cite{Sciortino:91} which in turn reduces the relative cost of
incorporating water molecules into a hydrophobic aggregate.  At least
two factors make water special in this context: its large negative
excess chemical potential prevents water from penetrating hydrophobic
aggregates under normal conditions,\cite{Hummer:PRL:98} thus driving
the formation of protein hydrophobic cores.  Under sufficiently high
pressure, however, the relatively small water molecules occupy
previously empty cavities, exploiting imperfections in the packing of
the protein interior.  Note that the small size of water molecules
here is not used as an argument that ``interstitial'' cavities are
smaller in water than in typical organic liquids,\cite{Lee:85,Lee:91}
as discussed in section \ref{sec:size}, but that water molecules can
occupy small cavities in the protein interior.  This incorporation of
water molecules is favored further by the versatility of water
molecules which can form favorable hydrogen-bonding interactions in
many different environments, particularly in the interior of a protein
which is never entirely nonpolar.

These arguments also explain in part the success of simple theories,
such as SPT and PC, that take no special account of the known details
of hydrophobic solvation patterning.  These theories yield
satisfactory predictions for the simplest hydrophobic effects in part
because they incorporate information about the equation of state and the
oxygen-atom pair correlations of water.  A next stage of basic molecular theory
of hydrophobic effects will surely address the hydration structure in
more detail while at the same time preserving equation of
state information such as the low, temperature insensitive values of
the isothermal compressibility that have been identified as
particularly important.

IT models of the solvation thermodynamics can easily
be generalized to various aqueous and non-aqueous solvents and
mixtures, combined with other approaches, or extended to ``unusual''
models of water, such as a recently proposed two dimensional model of
water\cite{Silverstein:JACS:98} or an isotropic water model without
directional hydrogen-bond interactions.\cite{Head-Gordon:JACS:95} We
have successfully adapted the method to study the solubility of small
molecules in polymeric fluids.\cite{Garde:MM:98,Cuthbert:MM:97} Crooks and
Chandler\cite{Crooks:PRE:97} compared solvation chemical potentials of
hard spheres of varying sizes in hard-sphere fluids to simulation
data, finding that a Gibbs prior gives a good description of the
density fluctuations and the solvation thermodynamics.  Applications
of the IT method to study phase equilibria have been
suggested.\cite{Wu:Prausnitz:98} Lum, Chandler, and
Weeks\cite{Lum:Chandler:Weeks:98} have combined field-theoretic
methods\cite{Chandler:PRE:93} with the IT approach to
describe the solvation thermodynamics and structure of mesoscopic and
macroscopic spherical solutes in water which is tightly associated
with ``dewetting'' of the solute
surface.\cite{Stillinger:73,Hummer:PRL:98}

The development of IT models is an ongoing effort
targeting several directions.  A better description for molecularly
large solutes can be expected from improved, physically motivated
default models that incorporate the multiphasic nature of water to
account for the free energy of forming microscopic vapor
bubbles.\cite{Lum:Chandler:Weeks:98} IT models can easily be
generalized to study hydrophobic interactions in inhomogeneous
environments, for instance, to describe ligand binding to proteins.
Section \ref{sec:cont} gives an outline of a continuous IT,
which establishes connections to the concepts of cavity
expulsion and dewetting of nonpolar surfaces.\cite{Hummer:PRL:98} The
volume stratification discussed in section \ref{sec:strat} leads
naturally to predictions of the hydration structure.

In conclusion, we believe that the basic IT model
provides a simple theoretical framework to study many hydrophobic
phenomena, as it already led to a new understanding of the temperature
dependence of hydrophobic hydration,\cite{Garde:PRL:96} and resulted
in a new description of the pressure denaturation of
proteins.\cite{Hummer:PNAS:98}

\acknowledgments

L.R.P. thanks Andrew Pohorille for extended collaborations on the
problems of hydrophobic effects.  This work was in part supported by
the U.S. Department of Energy through the Los Alamos National
Laboratory LDRD-CD grant for an ``Integrated Structural Biology
Resource.''  S.G. is a Director's Funded Postdoctoral Fellow at Los
Alamos National Laboratory.


\begin{thebibliography}{100}

\bibitem{Kauzmann:APC:59}
Kauzmann, W. {\em Adv. Protein Chem.} {\bf 1959}, {\em 14},  1.

\bibitem{Tanford:80}
Tanford, C.  {\em The Hydrophobic Effect: Formation of Micelles and Biological
  Membranes};  John Wiley {\&} Sons: New York, 1973.

\bibitem{Dill:B:90}
Dill, K.~A. {\em Biochemistry} {\bf 1990}, {\em 29},  7133.

\bibitem{Blokzijl:AC:93}
Blokzijl, W.; Engberts, J. B. F.~N. {\em Angew. Chem. Int. Ed. Engl.} {\bf
  1993}, {\em 32},  1545.

\bibitem{Fink:FD:98}
Fink, A.~L. {\em Folding {\&} Design} {\bf 1998}, {\em 3},  R9.

\bibitem{Privalov:APC:79}
Privalov, P.~L. {\em Adv. Prot. Chem.} {\bf 1979}, {\em 33},  167.

\bibitem{Privalov:APC:88}
Privalov, P.~L.; Gill, S.~J. {\em Adv. Prot. Chem.} {\bf 1988}, {\em 39},  191.

\bibitem{Makhatadze:APC:95}
Makhatadze, G.~I.; Privalov, P.~L. {\em Adv. Prot. Chem.} {\bf 1995}, {\em 47},
   307.

\bibitem{Richards:JMB:74}
Richards, F.~M. {\em J. Mol. Biol.} {\bf 1974}, {\em 82},  1.

\bibitem{Buckle:JMB:93}
Buckle, A.~M.; Henrick, K.; Fersht, A.~R. {\em J. Mol. Biol.} {\bf 1993}, {\em
  234},  847.

\bibitem{Baldwin:PNAS:86}
Baldwin, R.~L. {\em Proc. Natl. Acad. Sci. USA} {\bf 1986}, {\em 83},  8069.

\bibitem{Baldwin:PNAS:92}
Baldwin, R.~L.; Muller, N. {\em Proc. Natl. Acad. Sci. USA} {\bf 1992}, {\em
  89},  7110.

\bibitem{Kauzmann:87}
Kauzmann, W. {\em Nature (London)} {\bf 1987}, {\em 325},  763.

\bibitem{Garde:PRL:96}
Garde, S.; Hummer, G.; Garc{\'{\i}}a, A.~E.; Paulaitis, M.~E.; Pratt, L.~R.
  {\em Phys. Rev. Lett.} {\bf 1996}, {\em 77},  4966.

\bibitem{Hummer:PNAS:98}
Hummer, G.; Garde, S.; {Garc\'{\i}a}, A.~E.; Paulaitis, M.~E.; Pratt, L.~R.
  {\em Proc. Natl. Acad. Sci. USA} {\bf 1998}, {\em 95},  1552.

\bibitem{Privalov:JMB:74}
Privalov, P.~L.; Khechinashvili, N.~N. {\em J. Mol. Biol.} {\bf 1974}, {\em
  86},  665.

\bibitem{Makhatadze:95}
Makhatadze, G.~I.; Privalov, P.~L. {\em Adv. Protein Chem.} {\bf 1995}, {\em
  47},  307.

\bibitem{Privalov:JMB:93}
Privalov, P.~L.; Makhatadze, G.~I. {\em J. Mol. Biol.} {\bf 1993}, {\em 232},
  660.

\bibitem{Brandts:70}
Brandts, J.~F.; Oliveira, R.~J.; Westort, C. {\em Biochemistry} {\bf 1970},
  {\em 9},  1038.

\bibitem{Zipp:73}
Zipp, A.; Kauzmann, W. {\em Biochemistry} {\bf 1973}, {\em 12},  4217.

\bibitem{Heremans:82}
Heremans, K. {\em Annu. Rev. Biophys. Bioeng.} {\bf 1982}, {\em 11},  1.

\bibitem{Weber:83}
Weber, G.; Drickamer, H.~G. {\em Quart. Rev. Biophys.} {\bf 1983}, {\em 16},
  89.

\bibitem{Silva:93}
Silva, J.~L.; Weber, G. {\em Annu. Rev. Phys. Chem.} {\bf 1993}, {\em 44},  89.

\bibitem{Jonas:94}
Jonas, J.; Jonas, A. {\em Annu. Rev. Biophys. Biomol. Struct.} {\bf 1994}, {\em
  23},  287.

\bibitem{Royer:95}
Royer, C.~A. {\em Methods Enzymol.} {\bf 1995}, {\em 259},  357.

\bibitem{Silva:97}
Silva, J.~L.; Foguel, D.; {Da Poian}, A.~T.; Prevelige, P.~E. {\em Curr. Opin.
  Struct. Biol.} {\bf 1996}, {\em 6},  166.

\bibitem{Zhang:95}
Zhang, J.; Peng, X.; Jonas, A.; Jonas, J. {\em Biochemistry} {\bf 1995}, {\em
  34},  8631.

\bibitem{Panick:98}
Panick, G.; Malessa, R.; Winter, R.; Rapp, G.; Frye, K.~J.; Royer, C.~A. {\em
  J. Mol. Biol.} {\bf 1998}, {\em 275},  389.

\bibitem{Baldwin:BJ:96}
Baldwin, R.~L. {\em Biophys. J.} {\bf 1996}, {\em 71},  2056.

\bibitem{Frank:JCP:45}
Frank, H.~S.; Evans, M.~W. {\em J. Chem. Phys.} {\bf 1945}, {\em 13},  507.

\bibitem{Nemethy:JCP:62}
N\'{e}methy, G.; Scheraga, H.~A. {\em J. Chem. Phys.} {\bf 1962}, {\em 36},
  3382.

\bibitem{Ben-Naim:80}
Ben-Naim, A.  {\em Hydrophobic Interactions};  Plenum: New York, 1980.

\bibitem{Besseling:JPCB:97}
Besseling, N. A.~M.; Lyklema, J. {\em J. Phys. Chem. B} {\bf 1997}, {\em 101},
  7604.

\bibitem{Madan:JPCB:97}
Madan, B.; Sharp, K. {\em J. Phys. Chem. B} {\bf 1997}, {\em 101},  11237.

\bibitem{LitNote}
Note that it is not within the scope of this article to review the entire
  literature devoted to the hydrophobic effect. An electronic literature search
  produced over 28,000 publications that contain the word hydrophobic in the
  title or list of keywords since 1974!

\bibitem{Reiss:59}
Reiss, H.; Frisch, H.~L.; Lebowitz, J.~L. {\em J. Chem. Phys.} {\bf 1959}, {\em
  31},  369.

\bibitem{Pierotti:63}
Pierotti, R.~A. {\em J. Phys. Chem.} {\bf 1963}, {\em 67},  1840.

\bibitem{Stillinger:73}
Stillinger, F.~H. {\em J. Solut. Chem.} {\bf 1973}, {\em 2},  141.

\bibitem{Hermann:JPC:72}
Hermann, R.~B. {\em J. Phys. Chem.} {\bf 1972}, {\em 76},  2754.

\bibitem{Chothia:N:74}
Chothia, C. {\em Nature} {\bf 1974}, {\em 248},  338.

\bibitem{Oobatake:PBMB:93}
Oobatake, M.; Ooi, T. {\em Prog. Biophys. Mol. Biol.} {\bf 1993}, {\em 59},
  237.

\bibitem{Sharp:S:91}
Sharp, K.~A.; Nicholls, A.; Fine, R.~F.; Honig, B. {\em Science} {\bf 1991},
  {\em 252},  106.

\bibitem{Simonson:94}
Simonson, T.; Brunger, A.~T. {\em J. Phys. Chem.} {\bf 1994}, {\em 98},  4683.

\bibitem{Giesen:94}
Giesen, D.~J.; Cramer, C.~J.; Truhlar, D.~G. {\em J. Phys. Chem.} {\bf 1994},
  {\em 98},  4141.

\bibitem{Tunon:94}
Tunon, I.; Silla, E.; Pascualahuir, J.~L. {\em J. Phys. Chem.} {\bf 1994}, {\em
  98},  377.

\bibitem{Fukunishi:JPC:96}
Fukunishi, Y.; Suzuki, M. {\em J. Phys. Chem.} {\bf 1996}, {\em 100},  5634.

\bibitem{Pratt:JCP:77}
Pratt, L.~R.; Chandler, D. {\em J. Chem. Phys.} {\bf 1977}, {\em 67},  3683.

\bibitem{Lazaridis:92}
Lazaridis, T.; Paulaitis, M.~E. {\em J. Phys. Chem.} {\bf 1992}, {\em 96},
  3847.

\bibitem{Pangali:79}
Pangali, C.; Rao, M.; Berne, B.~J. {\em J. Chem. Phys.} {\bf 1979}, {\em 71},
  2975.

\bibitem{Hummer:PNAS:96}
Hummer, G.; Garde, S.; Garc{\'{\i}}a, A.~E.; Pohorille, A.; Pratt, L.~R. {\em
  Proc. Natl. Acad. Sci. USA} {\bf 1996}, {\em 93},  8951.

\bibitem{Jaynes:83} Jaynes, E.~T. In {\em E. T. Jaynes: Papers on
    Probability, Statistics, and Statistical Physics}; Rosenkrantz,
  R.~D., Ed.; Reidel: Dordrecht, Holland, 1983.

\bibitem{Chandler:PRE:93}
Chandler, D. {\em Phys. Rev. E} {\bf 1993}, {\em 48},  2898.

\bibitem{Berne:PNAS:96}
Berne, B.~J. {\em Proc. Natl. Acad. Sci. USA} {\bf 1996}, {\em 93},  8800.

\bibitem{Percus:JP:93}
Percus, J.~K. {\em J. Physique IV} {\bf 1993}, {\em 3},  49.

\bibitem{Hummer:PRL:98}
Hummer, G.; Garde, S. {\em Phys. Rev. Lett.} {\bf 1998}, {\em 80},  4193.

\bibitem{Pollack:S:91}
Pollack, G.~L. {\em Science} {\bf 1991}, {\em 251},  1323.

\bibitem{Berendsen:81} Berendsen, H. J.~C.; Postma, J. P.~M.; {van
    Gunsteren}, W.~F.; Hermans, J. In {\em Intermolecular Forces};
  Pullman, B., Ed.; Reidel: Dordrecht, Holland, 1981, pp.\ 331--342.

\bibitem{Hummer:ice:94}
Hummer, G.; Soumpasis, D.~M. {\em Phys. Rev. E} {\bf 1994}, {\em 49},  591.

\bibitem{Shore:80}
Shore, J.~E.; Johnson, R.~W. {\em IEEE Transactions on Information Theory} {\bf
  1980}, {\em 26},  26.

\bibitem{Berendsen:87}
Berendsen, H. J.~C.; Grigera, J.~R.; Straatsma, T.~P. {\em J. Phys. Chem.} {\bf
  1987}, {\em 91},  6269.

\bibitem{Pratt:NATO:98} Pratt, L.~R.; Garde, S.; Hummer, G.  {\em
    Proceedings of the NATO Advanced Study Institute: New Approaches
    to old and new Problems in Liquid State Theory, Patti Marina
    (Messina) Italy, July 1998}; submitted [LA-UR 98-98-2712].

\bibitem{Widom:63}
Widom, B. {\em J. Chem. Phys.} {\bf 1963}, {\em 39},  2808.

\bibitem{Stratonovich:JETP:55}
Stratonovich, R.~L. {\em J. Exper. Theoret. Phys. USSR} {\bf 1955}, {\em 28},
  409, english translation in {\em Soviet Physics-JETP} {\bf 1955}, {\em 1},
  254.

\bibitem{Hill:58}
Hill, T.~L. {\em J. Chem. Phys.} {\bf 1958}, {\em 28},  1179.

\bibitem{Guinier:39}
Guinier, A. {\em Ann. Phys. (Paris)} {\bf 1939}, {\em 1},  11.

\bibitem{Mayer:41}
Mayer, J.~E.; Montroll, E. {\em J. Chem. Phys.} {\bf 1941}, {\em 8},  2.

\bibitem{Riordan:78}
Riordan, J.  {\em An Introduction to Combinatorial Analysis};  Princeton
  University Press: Princeton, NJ, 1978, p.\ 32.

\bibitem{Reiss:65}
Reiss, H. {\em Adv. Chem. Phys.} {\bf 1965}, {\em 9},  1.

\bibitem{Pratt:PNAS:92}
Pratt, L.~R.; Pohorille, A. {\em Proc. Natl. Acad. Sci. USA} {\bf 1992}, {\em
  89},  2995.

\bibitem{Pohorille:JACS:90}
Pohorille, A.; Pratt, L.~R. {\em J. Am. Chem. Soc.} {\bf 1990}, {\em 112},
  5066.
  
\bibitem{Palma} Pratt, L.~R.; Pohorille, A. In {\em Proceedings of the
    {EBSA} (Association of the European Biophysical Societies) 1992
    International Workshop on Water-Biomolecule Interactions}; Palma,
  M.~U.; Palma-Vittorelli, M.~B.; Parak, F., Eds.; Societ\'{a}
  Italiana de Fisica: Bologna, 1993, pp.\ 261--268.

\bibitem{Guillot:JCP:93}
Guillot, B.; Guissani, Y. {\em J. Chem. Phys.} {\bf 1993}, {\em 99},  8075.

\bibitem{Forsman:94}
Forsman, J.; {J\"{o}nsson}, B. {\em J. Chem. Phys.} {\bf 1994}, {\em 101},
  5116.

\bibitem{Beutler:95:a}
Beutler, T.~C.; {B\'{e}guelin}, D.~R.; {van Gunsteren}, W.~F. {\em J. Chem.
  Phys.} {\bf 1995}, {\em 102},  3787.

\bibitem{Stamatopoulou:JCP:98}
Stamatopoulou, A.; Ben-Amotz, D. {\em J. Chem. Phys.} {\bf 1998}, {\em 108},
  7294.

\bibitem{Lee:85}
Lee, B. {\em Biopolymers} {\bf 1985}, {\em 24},  813.

\bibitem{Lee:91}
Lee, B. {\em Biopolymers} {\bf 1991}, {\em 31},  993.

\bibitem{Richards:SA:91}
Richards, F.~M. {\em Scientific American} {\bf 1991}, {\em 264},  54.

\bibitem{Haymet:Fara:96}
Haymet, A. D.~J.; Silverstein, K. A.~T.; Dill, K.~A. {\em Faraday Discussions}
  {\bf 1996}, {\em 103},  117.

\bibitem{Watanabe:86}
Watanabe, K.; Andersen, H.~C. {\em J. Phys. Chem.} {\bf 1986}, {\em 90},  795.

\bibitem{Smith:JCP:93}
Smith, D.~E.; Haymet, A. D.~J. {\em J. Chem. Phys.} {\bf 1993}, {\em 98},
  6445.

\bibitem{Wodak:93}
{van Belle}, D.; Wodak, S.~J. {\em J. Am. Chem. Soc.} {\bf 1993}, {\em 115},
  647.

\bibitem{Head-Gordon:JACS:95}
Head-Gordon, T. {\em J. Am. Chem. Soc.} {\bf 1995}, {\em 117},  501.

\bibitem{Garde:96:b}
Garde, S.; Hummer, G.; Paulaitis, M.~E. {\em Faraday Discussions} {\bf 1996},
  {\em 103},  125.

\bibitem{Ludemann:97}
{L\"{u}demann}, S.; Abseher, R.; Schreiber, H.; Steinhauser, O. {\em J. Am.
  Chem. Soc.} {\bf 1997}, {\em 119},  4206.

\bibitem{Hummer:96:a}
Hummer, G.; Pratt, L.~R.; {Garc\'{\i}a}, A.~E. {\em J. Phys. Chem.} {\bf 1996},
  {\em 100},  1206.

\bibitem{Jorgensen:84}
Jorgensen, W.~L.; Madura, J.~D.; Swenson, C.~J. {\em J. Am. Chem. Soc.} {\bf
  1984}, {\em 106},  6638.

\bibitem{Beglov:94}
Beglov, D.; Roux, B. {\em J. Chem. Phys.} {\bf 1994}, {\em 100},  9050.

\bibitem{Dill:S:90}
Dill, K.~A. {\em Science} {\bf 1990}, {\em 250},  297.

\bibitem{Rowlinson:Swinton:82}
Rowlinson, J.~S.; Swinton, F.~L.  {\em Liquids and Liquid Mixtures};
  Butterworths: London, 1982.

\bibitem{Sawamura:JPC:89}
Sawamura, S.; Kitamura, K.; Taniguchi, Y. {\em J. Phys. Chem.} {\bf 1989}, {\em
  93},  4931.

\bibitem{Payne:97}
Payne, V.~A.; Matubayasi, N.; Murphy, L.~R.; Levy, R.~M. {\em J. Phys. Chem. B}
  {\bf 1997}, {\em 101},  2054.

\bibitem{Wallqvist:JCP:92}
Wallqvist, A. {\em J. Chem. Phys.} {\bf 1992}, {\em 96},  1657.

\bibitem{Wallqvist:JPC:95:b}
Wallqvist, A.; Berne, B.~J. {\em J. Phys. Chem.} {\bf 1995}, {\em 99},  2893.

\bibitem{Wallqvist:JACS:98}
Wallqvist, A.; Covell, D.~G.; Thirumalai, D. {\em J. Am. Chem. Soc.} {\bf
  1998}, {\em 120},  427.

\bibitem{Sloan:90}
{Sloan, Jr.}, E.~D.  {\em Clathrate Hydrates of Natural Gases};  M. Dekker: New
  York, 1990.

\bibitem{Vidugiris:95}
Vidugiris, G. J.~A.; Markley, J.~L.; Royer, C.~A. {\em Biochemistry} {\bf
  1995}, {\em 34},  4909.

\bibitem{Bryngelson:95}
Bryngelson, J.~D.; Onuchic, J.~N.; Socci, N.~D.; Wolynes, P.~G. {\em Proteins
  Struct. Funct. Genet.} {\bf 1995}, {\em 21},  167.

\bibitem{Hummer:JPCM:94}
Hummer, G.; Soumpasis, D.~M.; {Neumann}, M. {\em J. Phys.: Condens. Matt.} {\bf
  1994}, {\em 6},  A141.

\bibitem{Garde:PRE:96}
Garde, S.; Hummer, G.; {Garc\'{\i}a}, A.~E.; Pratt, L.~R.; Paulaitis, M.~E.
  {\em Phys. Rev. E} {\bf 1996}, {\em 53},  R4310.

\bibitem{Ashbaugh:96}
Ashbaugh, H.~S.; Paulaitis, M.~E. {\em J. Phys. Chem.} {\bf 1996}, {\em 100},
  1900.

\bibitem{Weeks:71}
Weeks, J.~D.; Chandler, D.; Andersen, H.~C. {\em J. Chem. Phys.} {\bf 1971},
  {\em 54},  5237.

\bibitem{Head-Gordon:PNAS:95}
Head-Gordon, T. {\em Proc. Natl. Acad. Sci. USA} {\bf 1995}, {\em 92},  8308.

\bibitem{Stillinger:74}
Stillinger, F.~H.; Rahman, A. {\em J. Chem. Phys.} {\bf 1974}, {\em 61},  4973.

\bibitem{Sciortino:91}
Sciortino, F.; Geiger, A.; Stanley, H.~E. {\em Nature (London)} {\bf 1991},
  {\em 354},  218.

\bibitem{Silverstein:JACS:98}
Silverstein, K. A.~T.; Haymet, A. D.~J.; Dill, K.~A. {\em J. Am. Chem. Soc.}
  {\bf 1998}, {\em 120},  3166.

\bibitem{Garde:MM:98}
Garde, S.; Khare, R.; Hummer, G.; unpublished.

\bibitem{Cuthbert:MM:97}
Cuthbert, T.~R.; Wagner, N.~J.; Paulaitis, M.~E. {\em Macromolecules} {\bf
  1997}, {\em 30},  3058.

\bibitem{Crooks:PRE:97}
Crooks, G.~E.; Chandler, D. {\em Phys. Rev. E} {\bf 1997}, {\em 56},  4217.

\bibitem{Wu:Prausnitz:98}
Wu, J.~Z.; Prausnitz, J.~M. {\em Ind. Eng. Chem. Res.} {\bf 1998}, {\em 37},
  1634.

\bibitem{Lum:Chandler:Weeks:98}
Lum, K.; Chandler, D.; Weeks, J.~D.  {\bf 1998}, submitted.

\end{thebibliography}

\end{document}